\pdfoutput=1
\documentclass[12pt]{report}          
\usepackage[intlimits]{amsmath}
\usepackage{amsfonts,amssymb}
\DeclareSymbolFontAlphabet{\mathbb}{AMSb}
\usepackage{apalike}
\usepackage{float,subfigure}
\usepackage[bf]{caption2}       
\setcaptionmargin{0.5in}
\usepackage{fancyheadings,fancybox,ifthen}
\usepackage{bu_ece_thesis}
\usepackage{url}
\usepackage{lscape,afterpage}
\usepackage{xspace}
\usepackage[dvips]{graphicx}
\newcommand{\figref}[1]{Figure~\ref{#1}}

\newcommand{\be}{\begin{equation}}
\newcommand{\ee}{\end{equation}}
\newcommand{\bea}{\begin{eqnarray}}
\newcommand{\eea}{\end{eqnarray}}

\newcommand{\la}{\langle}
\newcommand{\ra}{\rangle}
\newcommand{\lb}{\left[}
\newcommand{\rb}{\right]}
\newcommand{\lp}{\left(}
\newcommand{\rp}{\right)}
\renewcommand{\vec}[1]{{\bf #1}}
\def\nn{\nonumber\\}

\begin{document}


\title{Cross-correlation (C$^2$) Imaging for Waveguide Characterization}

\author{Roman A. Barankov}

\degree=1

\prevdegrees{
Ph.D., Massachusetts Institute of Technology, 2006}

\department{Department of Electrical and Computer Engineering}

\defenseyear{2012}
\degreeyear{2012}

\reader{First}{Siddharth Ramachandran, Ph.D.}{Associate Professor of Electrical Engineering}
\reader{Second}{Jerome Mertz, Ph.D.}
{Professor of Biomedical Engineering}
\reader{Third}{M. Selim \"Unl\"u, Ph.D.}{Professor of Biomedical and Electrical Engineering}
\majorprof{Siddharth Ramachandran, Ph.D.}{\mbox{Department of Electrical and\\
    Computer Engineering}, \mbox{secondary appointment}}




\maketitle
\cleardoublepage

\copyrightpage
\cleardoublepage

\approvalpage
\cleardoublepage


 \newpage
 \section*{\centerline{Acknowledgments}}

First of all, I thank my wife Tatiana Barankova and my son Roman Barankov for
their continuous support of my scientific endeavors. We have been together
through a very difficult and interesting time. My scientific goals of these
two years have been accomplished thanks to their invaluable help.

With my background in theoretical condensed matter physics, it was not a
simple task to become an experimental physicist. I thank Prof. S. Ramachandran
for giving me this opportunity, introducing me to the exciting field of
applied optics, and for providing his support and the resources of his lab to
do very interesting and useful experiments. As a result, I have learned very
valuable skills and substantially extended my scientific background.

I also thank all the members of {\it Nanostructured Fibers and Nonlinear
Optics} laboratory for their good company during my stressful and creative
journey in Optics. The hard work combined with deep interest in science I
witnessed in the laboratory have constantly stimulated my research during
these two years.
 
 \cleardoublepage


\begin{abstractpage}


Confined geometries, such as optical waveguides, support a discrete set of
eigen-modes. In multimoded structures, depending on the boundary conditions,
superposition states can propagate. Characterization of these states is a
fundamental problem important in waveguide design and testing, especially for
optical applications.

In this work, I have developed a novel interferometric method that provides
complete characterization of optical waveguide modes and their superposition
states. The basic idea of the method is to study the interference of the beam
radiated from an optical waveguide with an external reference beam, and detect
different waveguide modes in the time-domain by changing the relative optical
paths of the two beams.

In particular, this method, called cross-correlation or C$^2$-imaging,
provides the relative amplitudes of the modes and their group delays. For
every mode, one can determine the dispersion, intensity and phase
distributions, and also local polarization properties.

As a part of this work, I have developed the mathematical formalism of
C$^2$-imaging and built an experimental setup implementing the idea. I have
carried out an extensive program of experiments, confirming the ability of the
method to completely characterize waveguide properties.

\end{abstractpage}
\cleardoublepage


\tableofcontents
\cleardoublepage


\newpage
\listoffigures
\cleardoublepage



\newpage

\cleardoublepage

\pagenumbering{arabic}

\chapter{Introduction}
\label{chapter:Introduction}
\thispagestyle{myheadings}

In optical applications, the demand for reliable waveguide characterization
methods comes from the development of several fiber-based optical
technologies. First, the network traffic in optical fiber communication
systems demands further increase in optical channels to overcome the
limitations of currently deployed wave-division multiplexing (WDM) systems,
such as the bandwidth of optical amplifiers and input power.

Mode-division multiplexing (MDM) pioneered in Ref.~\cite{Berdague1982} and
recently developed in Refs.\cite{Hanzawa2011,Ryf2011,Salsi2011} is one
possible solution of the network-capacity problem. MDM systems employing
several fiber modes for transmission of information require real-time
monitoring of modal power decomposition.

Several approaches have been suggested to realize the monitoring, such as the
far-field imaging of the output facet of multimoded fibers~\cite{Rittich1985},
and a method using ultrafast sources and specially designed probe fibers to
temporally resolve the launched power~\cite{Golowich2004}. A promising
alternative for modal power monitoring in real-time is provided by the
tilted-fiber Bragg gratings. The gratings couple guided modes to radiation
modes encoding, among other characteristics, the modal power distribution in
the probed
systems~\cite{Wagener1997,Westbrook2000,Feder2003,Yang2005,Yan2012}.

High-performance, high-power fiber-lasers~\cite{Richardson2010} is another
area in which waveguide characterization is of outmost importance. The design
of novel fiber-platforms capable of achieving excellent output beam-qualities
requires the propagation of only one mode in fibers that are not strictly
single-moded~\cite{Dong2009,Ramachandran2006,Galvanauskas2008,Stutzki2011,Koplow2000}.
A usual measure of the laser-beam quality in these systems is the so-called
M$^2$-parameter~\cite{Siegman1990,Siegman1993}. However, being essentially an
integral characteristic, it gives a very rough estimate of higher-order mode
content~\cite{Wielandy2007}.

A recently reported characterization technique -- S$^2$
imaging~\cite{Nicholson2008,Nicholson2009} -- enables the direct measurement
of modal content of multimoded waveguides by recording spatially resolved
output of the fiber in frequency domain. Historically, this approach appeared
from the early studies of mode dispersion of higher-order
modes~\cite{Menashe2001} and multi-path
interference~\cite{Ramachandran2003,Ramachandran2005_MPI,Ramachandran2005}. In
S$^2$-imaging method, one of the modes, usually the fundamental one, is used
as a ``reference" mode for the analysis of the interference signal, which is a
fundamental impediment of the approach. In particular, in this case, a
reasonable estimate of the relative power levels of the modes may be obtained
only when the ``reference'' mode has the largest power in the propagating
beam. Moreover, the presence of multiple interferences of every mode with all
other modes complicates the analysis of the recorded signal and makes the
power reconstruction unreliable. As a result, the method fails to characterize
the generic case of multiple modes having similar power levels. Yet another
difficulty is related to the inherent insensitivity of the method to the
polarization properties of the modes, which introduces uncontrollable
approximations in the interpretation of the results, in the general case of
elliptically polarized states.

An alternative approach of optical low-coherence interferometry does not
suffer from the limitations of S$^2$-imaging, since it employs the external
reference beam with well-defined characteristics, which interferes with all
the modes of the test waveguide~\cite{Nandi2009,Ma2009}. This technique does
not require any prior knowledge of group-delay or dispersive characteristics
of the reference mode. In contrast to S$^2$ imaging, all the modes having
arbitrary relative power levels and polarization properties can be measured
independently of one another.

In this work, I have developed and realized this idea. In particular, I have
worked out the mathematical formalism forming the basis of the method and
demonstrated that interferometric signal contains complete information about
waveguide modes, including modal decomposition, intensity and phase
distributions of all the modes, and also their polarization properties. The
formalism was applied for characterization of several optical waveguides with
distinct properties and successfully demonstrated all the declared
capabilities.

The core of my Thesis begins in Chapter~\ref{chapter:Formalism} with general
discussion of the mathematical formalism of cross-correlation imaging.
Application of the method to characterization of several optical waveguides is
demonstrated in Chapter~\ref{chapter:Experiments}. The summary of my
experimental and theoretical work is presented in
Chapter~\ref{chapter:Conclusion}.

\cleardoublepage

\chapter{Mathematical formalism of C$^2$-imaging}
\label{chapter:Formalism}
\thispagestyle{myheadings}

In this Chapter, I discuss the mathematical formalism of
C$^2$-imaging~\cite{Barankov2012_Review}, which forms the basis of the
experimental method applied in Chapter~\ref{chapter:Experiments} for
characterization of several optical waveguides. In
Section~\ref{section:C2_General}, I introduce the general framework of
waveguide interferometry necessary for basic applications of the method. Then,
in Section~\ref{section:C2_Polarization}, I present a modification of the
method suitable for studies of polarization properties of waveguide modes.
Finally, in Section~\ref{section:C2_Phase}, I describe phase-sensitive
modification of the method.

\section{General description of C$^2$-imaging}
\label{section:C2_General}

In this section, I provide the basic description of the mathematical formalism
of C$^2$-imaging method. Some elements of the formalism were already presented
in Ref.~\cite{Schimpf2011}; here, I expand and further develop this early
discussion.

In multimoded optical waveguides, a beam of light propagates as a
superposition of discrete modes characterized by different propagation
constants and various intensity patterns of the modes, depending on the
boundary conditions. Direct imaging of the beam provides the intensity
distribution and also the output power averaged over superposition of all the
modes. The goal of this section is to demonstrate a novel interferometric
approach that allows characterization of modal powers and intensity
distribution of every waveguide mode contributing to the superposition state.

The basic idea of the method is to study the interference of the beam radiated
from an optical waveguide with an external reference beam and detect
different waveguide modes in the time-domain by changing the relative optical
paths of the two beams.

\subsection{Interferometry of optical beams}

\begin{figure}[!htb]
 	\includegraphics[width=12cm]{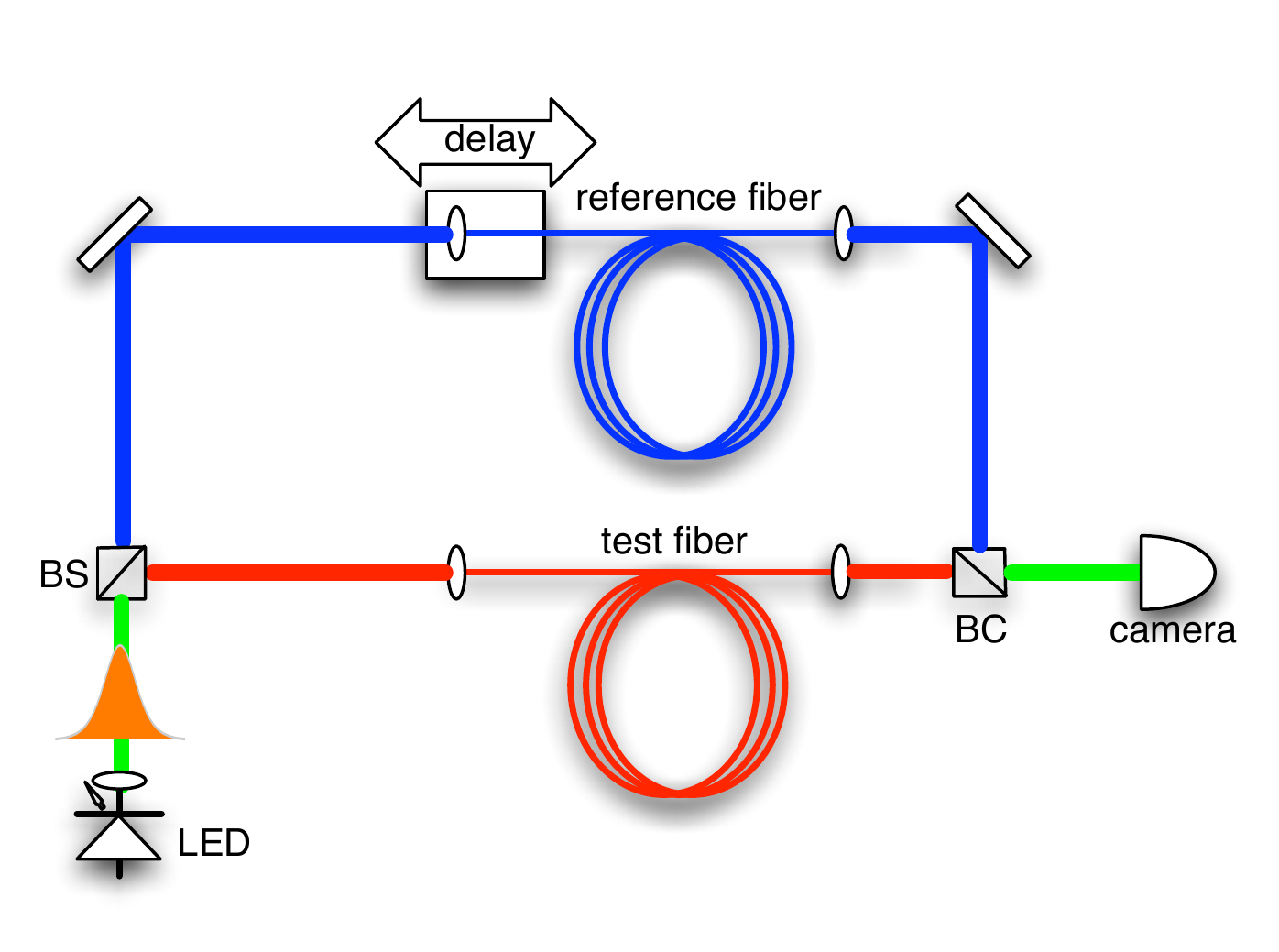}
  \caption{C$^2$-imaging setup: LED -- light-emitting diode, BS -- beam-splitter, BC -- beam combiner, Delay -- delay stage.}
  \label{fig:setup_basic}
\end{figure}

In this work, we employ the standard Mach-Zehnder interferometer configuration
shown in \figref{fig:setup_basic} to study the interference of the reference
and the test beam radiated from the signal waveguide. The beam of light from
an LED source is divided into two beams at the beam-splitter and then, after
propagating in the reference and test waveguides, recombined at the beam
combiner. The beam of light, radiated from a single-moded reference fiber, is
collimated, while the signal beam of the test fiber is focused at the camera.
The camera records a stack of images at different positions of the delay
stage.

The electric field at the image plane of the camera is a superposition of two
fields
\be\label{eq:total_field}
\vec E(\vec r,t)=\vec E_r(\vec r,t)+\vec E_s(\vec r,t+\tau),
\ee
where $\vec E_r(\vec r,t)$ and $\vec E_s(\vec r,t)$ are the electric fields of
the reference and signal beams, correspondingly.

The signal beam is delayed by time 
\be
\tau=d/c
\ee
with respect to the reference beam, where $d$ is the free-space difference
between the two optical paths, and $c$ is the speed of light in vacuum. The
time delay is introduced by the delay stage shown in
Fig.~\ref{fig:setup_basic}.

The camera records the intensity of light averaged over the exposure time,
which allows to employ the two equivalent representations of the electric
fields in time and frequency domains, $\vec E(\vec r,\omega)=\int dt\,
e^{i\omega t}\vec E(\vec r,t)$, according to the well-known identity
(Parseval's theorem):
\be\label{eq:Parseval} 
I(\vec r,\tau)= \int \limits_{- \Delta T/2}^{+\Delta T/2} \mbox{d} t  |\vec E(\vec r,t)|^2 = \int \limits_{-\infty}^{+\infty}\frac{ \mbox{d} \omega}{2\pi}\, | \mathbf{ E}(\vec r ,\omega)|^2,
\ee
where $\vec r=(x,y)$ defines the position of the camera pixel in the imaging
plane, and $\Delta T$ is the exposure time of the camera, assumed to be much
larger than the characteristic period of the electric field oscillations.

Upon substitution of the electric field~(\ref{eq:total_field}) into
Eq.~(\ref{eq:Parseval}), we arrive at the expression for the average intensity
that contains the background intensities of the reference and the signal beams
$I_0$, independent of the time-delay $\tau$, and also the interferometric
intensity $I_{int}$
\be
I(\vec r,\tau)=I_0(\vec r)+I_{int}(\vec r,\tau),
\ee
with two terms written explicitly as
\be\label{eq:back_intensity}
I_0(\vec r)=\int_{-\infty}^{+\infty}\frac{d\omega}{2\pi}\, \lp|\vec E_r(\vec r,\omega)|^2+|\vec E_s(\vec r,\omega)|^2\rp,
\ee
\bea\label{eq:int_intensity}
I_{int}(\vec r,\tau)&=&2\mathbf{Re}\,\int_{-\infty}^{\infty}dt\,\vec E^*_r(\vec r,t)\vec E_s(\vec r,t+\tau)\nn
&=& 2
{\rm\bf Re}\,\int_{-\infty}^{+\infty}\frac{d\omega}{2\pi}\, \vec
E_r^*(\vec r,\omega)\vec E_s(\vec r,\omega)e^{-i\omega\tau},
\eea
According to these equations, the interferometric signal $I_{int}\sim\vec
E_r^*\vec E_s\sim e^{i\varphi_{rs}-i\omega\tau}$ is sensitive to the optical
path difference $\varphi_{rs}=\varphi_r-\varphi_s$ and also the time-delay
$\tau$.

\subsection{Analysis of interferometric signal}

The electric fields of the reference and signal beams in
Eqs.~(\ref{eq:back_intensity}) and~(\ref{eq:int_intensity}) are given by
\bea\label{eq:fields} 
\vec E_r(\vec r,\omega)&=&\vec e_r(\vec r,\omega)A_r(\omega)e^{i\varphi_r},\nn
\vec E_s(\vec r,\omega)&=&\sum_m \alpha_m \vec e_m(\vec r,\omega)A_m(\omega)e^{i\varphi_m},
\eea
where the electric field $\vec E_s$ is a superposition of waveguide modes
$\vec E_m$ with real modal amplitudes $\alpha_m$. The power of
the $m$-th mode is given by
\be
p_m=\alpha_m^2.
\ee
The spectral properties of the reference beam and waveguide modes are encoded
in the corresponding functions $A_r\lp\omega\rp$ and $A_m\lp\omega\rp$.

Every mode of the signal beam and the reference beam is characterized by the intensity distribution:
\be
I_{ref}(\vec r,\omega)=|\vec e_{r}(\vec r,\omega)|^2,\, I_{m}(\vec r,\omega)=|\vec e_{m}(\vec r,\omega)|^2.
\ee
The intensity distribution $I_{ref}(\vec r,\omega)$ of the reference beam is
imaged directly by blocking the signal path of the interferometer. A similar
direct measurement of the signal beam, done by blocking the reference path,
produces only the total intensity distribution $I_{tot}=\sum_m I_m$ of all the
modes propagating in the signal beam, and, thus, does not provide access to
the intensities of individual modes.

The goal of this section is to demonstrate that C$^2$-imaging allows to
separately measure the intensity distribution and relative power level of
every mode of the signal beam, employing the property of the modes to
experience different phase shifts upon propagation in a waveguide.

Specifically, the $m$-th mode, after propagation in the test fiber of length $L$, acquires the phase
\be\label{eq:ph_m}
\varphi_m = \beta_m\left(\omega\right) L,
\ee
where $\beta_m\lp\omega\rp$ is the propagation constant of the mode.

The phase of the reference field, propagating in a single-mode reference fiber
of length $L_r$, is given by
\be\label{eq:ph_ref}
\varphi_r = \beta_r\left(\omega \right)  L_r,
\ee
where $\beta_r(\omega)$ is the propagation constant, and $L_r$ is the length
of the reference fiber.

The transverse phase-profile of the reference beam is assumed flat at the
imaging plane, which is satisfied by the collimation of the beam and also
ensuring that the imaging plane coincides with the neck of the collimated
beam. Regarding the signal beam, the near-field image of the output facet of
the signal fiber is recorded, by focusing the beam at the imaging plane.

By substituting the electric field of Eq.~(\ref{eq:fields}) into Eq.~(\ref{eq:int_intensity}), we arrive at the expression for the intensity
$I_{int}(\vec r,\tau)$, which accounts for the interference between the
reference field and the individual modes:
\be \label{eq:int}
I_{int}(\vec r,\tau)= 2\textbf{Re}\sum_m \int\limits_{-\infty}^{+\infty}\frac{\mbox{d} \omega}{2\pi}\, \vec e_{r}^*(\vec r,\omega) A_{r}^*(\omega)  \alpha_{m} \vec e_{m}(\vec r,\omega)A_{m}(\omega)e^{i(\varphi_m-\varphi_r-\omega\tau)}.
\ee
The analysis of this expression simplifies if one makes several reasonable
assumptions. In particular, since we are mainly interested in optical
measurements, typically performed with relatively narrow spectral widths
around the central frequency of the source, it is usually safe to assume that
the electric fields do not significantly vary in this spectral window
\be
\mathbf{e}\left(
\vec r,\omega \right) \approx \mathbf{e}\left( \vec r,\omega _0\right),
\ee
where $\omega_0$ is the central frequency of the light source.

Then Eq.~(\ref{eq:int}) simplifies:
\be \label{eq:interference}
I_{int}(\vec r,\tau)=2 \textbf{Re}\sum_m \alpha_m\vec e_{r}^*(\vec r,\omega_0)\vec e_{m}(\vec r,\omega_0) G_{rm} \left( \tau + \tau_{rm} \right)e^{-i\theta_{rm}}, 
\ee
where I introduced the joint coherence function
\bea\label{eq:jcf}
G_{rm}\left(t\right)&=&\int \frac{d\Omega}{2\pi}\,e^{-i \Omega t}\, S_{rm}\left( \Omega \right),\nn
S_{rm}(\Omega)&=&A^*_r(\Omega)A_m(\Omega)e^{-i\varphi_{rm}\left(\Omega \right)},
\eea
and defined the spectral functions with respect to a shifted frequency
\be
\Omega=\omega-\omega_0.
\ee

The joint coherence function depends on the differential group-delay of every
mode propagating in the signal beam with respect to the reference beam
\be
\tau_{mr} =L/v_{gr,m}-L_r/v_{gr,r}.
\ee

The group-velocity dispersion mismatch between the two beams is accounted for by the frequency-dependent phase
\be
\varphi_{mr}=\sum_{k\ge2}(\beta_m^{(k)}L-\beta_r^{(k)}L_r)\Omega^k/k!,
\ee 
where $\beta^{(k)}$ stand for the Taylor-coefficients of the mode-propagation constant $\beta$ calculated at the central frequency $\omega_0$ of the source.

The dominant contribution to the time-delay variation of the interferometric
intensity comes from the zero-order optical path mismatch of the beams
defined at the central frequency of the source
\be
\theta_{mr}=\beta_m^{(0)}L-\beta_r^{(0)}L_r-\omega_0\tau.
\ee

The joint coherence function describes the impact of the frequency-dependent
mode-propagation constant and the spectral properties of the fields on the
interferometric signal. In particular, the group-delay $\tau_{mr}$ defines the
position of $m$-th mode in the interferometric trace recorded by the camera at
a given transverse position in the imaging plane, and the shape of the trace
is determined by the joint spectral function of the reference and the signal
beams $S_{rm}$. In cases when waveguide modes do not undergo spectral
filtering (e.g. due to a long period grating embedded in the waveguide), the
spectral functions of the two beams are identical, $A=A_m=A_r$, and the joint
spectral function is given by $S_{rm}=S_0e^{-i\varphi_{rm}}$, where
$S_0=|A|^2$ is the spectrum of the light source.

The general expressions~(\ref{eq:interference}) and~(\ref{eq:jcf}), describing
interference of the signal and the reference beams, form the basis of further
theoretical analysis and experiments.

\subsection{Determination of mode dispersion and relative group delays}

The structure of the coherence function defines dispersive properties of the
modes. Specifically, the complex-valued joint coherence function may be
represented as a product of the real-valued amplitude and the phase factor:
\be
G_{mr}(\tau)={\cal G}_{mr}(\tau)e^{-i\psi_{mr}(\tau)},
\ee
which certainly depend on the detailed structure of the source spectrum. In
general, however, these are slowly varying functions on the time scale of the
electric field oscillations $2\pi/\omega_0$. 

An important property of the joint coherence function is the normalization,
which is independent of the dispersive properties of the modes. Indeed, it is
determined only by the spectral properties, as it follows from the identities
\be
\int\limits_{-\infty}^{\infty}d\tau\,
|G_{mr}(\tau)|^2=\int\limits_{-\infty}^{\infty}d\tau\,{\cal G}^2_{mr}(\tau)=\int\limits_{-\infty}^{\infty}
\frac{d\Omega}{2\pi}\,\left|A_m(\Omega)\right|^2\left|A_r(\Omega)\right|^2.
\ee

Normalization condition also indicates that the amplitude of the joint
coherence function ${\cal G}_{mr}(t)$ has finite extent $\Lambda_m$ in the
time domain defined as
\be
t\in\Lambda_m: |t|\le\delta t_m\sim 1/\Delta\omega,
\ee
which is dictated by the smallest spectral width of the reference and the
signal beams $\Delta\omega$. The time-extent $\delta t_m$ is usually
mode-dependent, due to the dispersion effects.

The separation of time-scales, $\delta t_m\gg 1/\omega_0$ suggests
time-averaging of the {\it square} of the interference
term~(\ref{eq:interference}) on the time-scale $1/\omega_0\ll\Delta t\ll\delta
t_m$. After the averaging, we obtain
\be\label{eq:P_r}
{\cal P}_r(\vec r, \tau)=\frac{1}{2\Delta t}\int\limits_{\tau-\Delta
t/2}^{\tau+\Delta t/2}dt\,I_{int}^2(\vec r,t)/I_{ref}(\vec r)=\sum_m p_m|\hat
e^*_r\hat e_m|^2 {\cal G}_{mr}^2(\tau-\tau_{mr}) I_{m}(\vec r),
\ee
where $p_m$ is the modal power, $\hat e_r$ and $\hat e_m$ are the
polarization unit vectors of the reference and signal beams, and
\be\label{eq:I_ref_I_m}
I_{ref}(\vec r)=|\vec e_{r}(\vec r,\omega_0)|^2,\, I_{m}(\vec r)=|\vec
e_{m}(\vec r,\omega_0)|^2
\ee
are the intensity distributions of the reference and signal beam at the
central frequency of the source.

By measuring the interference signal for a full set of polarization states of
the reference beam, one finds the intensity as a function of the time-delay:
\be\label{eq:P}
{\cal P}(\vec r, \tau)=\sum_r {\cal P}_{r}(\vec r, \tau)=\sum_m p_m{\cal G}_{mr}^2(\tau-\tau_{mr}) I_{m}(\vec r).
\ee

This expression indicates a strategy for separating waveguide modes in the
time-domain. In particular, provided the differential group delay of the two
nearby modes is larger than the characteristic extent of the coherence
function, i.e. $\tau_{m,m+1}\gtrsim \delta t$, the interference peaks
corresponding to these modes are well-separated in the time-domain. In this
case, the intensity ${\cal P}(\vec r, \tau)$ at every position $\vec r$ in the
imaging plane exhibits a series of well-separated peaks located at
$\tau=\tau_{mr}$, having the shapes determined by the spectral functions of
the modes.

Therefore, by employing an appropriate fitting procedure of the joint
coherence function to the measured interferometric signal, one finds the
relative group delays and determines dispersive properties of the modes. In
cases when the dispersion of the reference beam is unknown, one can measure
the interferometric signal at several different lengths of the reference
waveguide, and then use the resulting set of equations to determine dispersion
characteristics of all the modes, including the reference mode.

\subsection{Reconstruction of modal weights and intensity distributions}

The structure of the joint coherence function allows determination of the
modal weights without detailed knowledge of the source spectrum. Indeed, the
modal intensity distribution and power of every mode can be found by, first,
integrating Eq.~(\ref{eq:P}) over the extent $\Lambda_m$ and then integrating
over the imaging plane:
\bea
I_m(\vec r) &=&\int\limits_{\Lambda_m} d\tau\,{\cal P}(\vec r,\tau),\nn
p_m &=&\int\limits_{\Lambda_m} d\tau\int d\vec r\,{\cal P}(\vec r,\tau),
\eea
where it is assumed that the modal intensity and spectral function (same for all modes) are normalized to unity, i.e.
\be
\int \frac{d\Omega}{2\pi} |S_{mr}(\Omega)|^2=1,\, \int d\vec r\, I_{m}(\vec r)=1.
\ee

The relative powers $p_m$ of all waveguide modes can be reliably extracted
from the interference signal only when the coherence peaks do not
significantly overlap. When some peaks do overlap, e.g. due to their
relatively small differential group delay, the outlined procedure may still be
used. However, in this case it will provide the relative power of the
corresponding group of overlapping peaks. The formalism presented above can
not discriminate the degenerate modes. Below, I introduce a
polarization-sensitive modification of the method that allows the
reconstruction of degenerate modes, using their polarization properties. One
example of such reconstruction is demonstrated in
Chapter~\ref{chapter:Experiments}. Other modifications, accounting for
mode-dependent spectral properties, are also possible.

\subsection{Gaussian model}
The general description of the beam interferometry presented in the previous
section is illustrated in this part, assuming same spectral characteristics of
the signal and reference beams, $A=A_m=A_r$, and using a gaussian model of the
light source
\be
S(\Omega)=S_0\exp{\lb-(\Omega/\Delta\Omega)^2\rb},\, \Omega=\omega-\omega_0,
\ee
centered at $\omega_0$ with the spectral width $\Delta\Omega$. To simplify the
calculation, I consider only the effects of group-velocity dispersion
\be
\varphi_{mr}\approx\lp\beta_m^{(2)}L-\beta_r^{(2)}L_r\rp\Omega^2/2.
\ee

Under these conditions, the integral in Eq.(\ref{eq:jcf}) can be analytically
calculated, which leads to the following expression for the interferometric
intensity
\be\label{eq:int_gauss}
{\cal P}(\vec r,\tau)=\sum_m  p_m I_m(\vec r) \frac{\Delta\Omega}{\sqrt{2\pi\lp 1+d_{mr}^2\rp}}\exp\lb -\frac{\lp\tau-\tau_{mr}\rp^2\Delta\Omega^2}{2\lp 1+d_{mr}^2\rp}\rb
\ee
where the dispersion effects are included into the dimensionless parameter
\be
d_{mr}=(\beta_m^{(2)} L - \beta_r^{(2)} L_r)  \Delta\Omega^2/2,
\ee
and the modal intensities $I_m(\vec r)$ are normalized.

Eq.~(\ref{eq:int_gauss}) indicates that relative height of the interference
peaks strongly depends on the dispersion effects: for larger dispersion, the
time extent of the peak increases while its height correspondingly decreases,
so that the normalization of the coherence function is preserved. This
observation is generally true for any spectrum properties of the reference and
signal beams and suggest the use of the integral characteristics instead of
the peak value for measurements of the modal powers.

\subsection{Dispersion compensation}\label{subsection:dispersion_compensation}

The temporal extent of the interference peaks corresponding to different modes
defines the temporal resolution of C$^2$-imaging. Most clearly this limitation
appears in the analysis of the gaussian model, where the temporal full-width
at half-maximum of the interference peak
\be\label{eq:resolution}
\Delta\tau_{FWHM}=\sqrt{8\ln 2}\sqrt{1+d^2_{mr}}/\Delta\Omega. 
\ee
is determined by the spectral width $\Delta\Omega$ of the source and
group-velocity dispersion mismatch between the two beams $d_{mr}$. Obviously,
this expression provides an estimate for the temporal resolution of the
method, assuming that a pair of the modes with small relative group delay have
similar group-velocity dispersions.

As it follows from Eq.~(\ref{eq:resolution}), temporal resolution approaches
the limiting value
\be\label{eq:resolution_limit}
\Delta\tau_{FWHM}=\sqrt{8\ln 2}/\Delta\Omega,
\ee
when the lengths of the reference and the test fibers are matched, i.e.
\be\label{eq:disp_compensation}
L_{opt}=L\,\beta_m^{(2)}/\beta_r^{(2)},
\ee
so that the group-velocity dispersion is fully compensated.

For an arbitrary non-gaussian spectrum, the condition of the
dispersion-compensation remains the same. Indeed, the dispersion effects are
accounted for by the frequency-dependent phase
\be
\varphi_{mr}= \sum_{k\ge2}\lp\beta_m^{(k)}L-\beta_r^{(k)}L_r\rp\Omega^k/k!,
\ee 
The structure of the phase shows that the impact of
dispersion is minimized when the length of the reference waveguide satisfies
\be\label{eq:disp_compensation_gen}
L_{opt}=L\,\beta_m^{(k)}/\beta_r^{(k)},
\ee
where $k$ is the order of the first non-vanishing dispersion term. The
dispersion effects usually become noticeable already at $k=2$, and, as
expected, this condition then reproduces Eq.~(\ref{eq:disp_compensation}).
It's worth noting that conditions~(\ref{eq:disp_compensation}),
and~(\ref{eq:disp_compensation_gen}) can be satisfied only when the ratios of
the corresponding dispersive coefficients $\beta_m^{(k)}/\beta_r^{(k)}>0$.

In practice, it is possible to compensate the dispersion effects for a single
mode or a group of modes with similar dispersion characteristics. The residual
dispersion of other non-compensated modes results in their temporal
broadening.

Certainly, the limit of the temporal resolution~(\ref{eq:resolution_limit})
can only be achieved for purely parabolic dispersion. In more general
situations, higher-order dispersive terms may become important. For example,
when the group-velocity dispersion is fully compensated, the third-order phase
term contributes a correction to the spectrally limited temporal resolution of
the order
\be
d^{(3)}_{mr}\simeq
L\Delta\Omega^3\lp\beta_r^{(2)}\beta_m^{(3)}-\beta_r^{(3)}\beta_m^{(2)}\rp/\lp6\beta_r^{(2)}\rp,
\ee
which depends on the spectral properties of the system and the length of the
test waveguide. Under usual conditions, this correction is expected to be
small.

\section{Polarization-sensitive imaging}\label{subsection:polarization}
\label{section:C2_Polarization}

In this section, I describe how C$^2$-imaging can be used to reconstruct
polarization state of the modes propagating in few-moded optical
waveguides~\cite{Barankov2012_Polarization}. Specifically, I show that the
measurement of the cross-correlation signal for six different polarization
states of the reference beam allows full characterization of the polarization
state of every mode of the test beam in terms of the spatially-dependent
Stokes parameters. A specific implementation and application of this method is
demonstrated in Chapter~\ref{chapter:Experiments}.

\subsection{Polarization states}
In spinor notations, every polarization state of $m$-th mode can be expressed as a vector with two components
\be
\hat e_m=\left(
\begin{array}{c}
 E_x \\
 E_y
\end{array}
\right)=e^{i\alpha_x}\left(
\begin{array}{c}
 |E_x| \\
 |E_y|e^{i\varphi}
\end{array}
\right),
\ee
where $\varphi=\alpha_y-\alpha_x$ is the relative phase of the complex
amplitudes $E_x=|E_x|e^{i\alpha_x}$ and $E_y=|E_y|e^{i\alpha_y}$. The absolute
phase, without loss of generality, can be neglected.

The complete characterization of the polarization state requires
identification of the following three pairs of orthonormal basis states. The
first pair is formed by the states linearly polarized along $x$- and $y$-axis:
\be
|H\ra=\left(
\begin{array}{c}
 1 \\
 0
\end{array}
\right),
\quad |V\ra=\left(
\begin{array}{c}
 0 \\
 1
\end{array}
\right).
\ee

Another pair is obtained by rotating the basis by $\pi/4$ radians:
\be
|+\ra=\frac{1}{\sqrt{2}}\left(
\begin{array}{c}
 1 \\
 1
\end{array}
\right),
\quad
|-\ra=\frac{1}{\sqrt{2}}\left(
\begin{array}{c}
 1 \\
 -1
\end{array}
\right).
\ee

The last pair is given by the left and right circularly polarized states:
\be
|L\ra=\frac{1}{\sqrt{2}}\left(
\begin{array}{c}
 1 \\
 -i
\end{array}
\right),
\quad
|R\ra=\frac{1}{\sqrt{2}}\left(
\begin{array}{c}
 1 \\
 i
\end{array}
\right).
\ee

Experimentally, projection onto these six states can be realized using
combinations of a linear polarizer, half-wave and also quarter-wave plate, characterized by the following matrices
\bea
T_{LP}(\theta)&=&
\lb
\begin{array}{cc}
 \cos^2\theta & \sin\theta\cos\theta \\
  \sin\theta\cos\theta & \sin^2\theta
\end{array}
\rb,\nn
T_{HWP}(\theta)&=&
\lb
\begin{array}{cc}
 -\cos 2\theta & -\sin 2\theta\\
 -\sin 2\theta & \cos 2\theta
\end{array}
\rb,\nn
T_{QWP}(\theta)&=&
\lb
\begin{array}{cc}
 i \cos^2\theta+\sin^2\theta & (i-1) \sin\theta\cos\theta \\
 (i-1) \sin\theta\cos\theta & i \sin^2\theta+\cos^2\theta
\end{array}
\rb,
\eea
where $\theta$ determines the angle of the fast axis of the corresponding polarizing element.

A specific implementation of the projections onto these states is demonstrated
in Chapter~\ref{chapter:Experiments}.

\subsection{Stokes parameters}
An arbitrary state of polarized or unpolarized light is characterized by the
Stokes vector $\vec S=(S_0,S_1,S_2,S_3)$ with the following components:
\bea
S_0&=&\la |E_x|^2\ra+\la |E_y|^2\ra,\nn
S_1&=&\la |E_x|^2\ra-\la |E_y|^2\ra,\nn
S_2&=&2\la |E_x E_y|\cos\varphi\ra,\nn
S_3&=&2\la |E_x E_y|\sin\varphi\ra,
\eea
where averaging $\la...\ra$ is taken with respect to an ensemble of
measurements. When $E_x$, $E_y$ and $\varphi$ do not vary within the ensemble,
the light is completely polarized.

\subsection{Polarization distribution of waveguide modes}

In C$^2$-imaging, the intensity distribution function is proportional
to the scalar product of the polarizations of the reference beam $\hat e_r$
and the $m$-th mode $\hat e_m(\vec r)=(E_x(\vec r),E_y(\vec r))^{T}$ (transposed vector) of the signal
beam:
\be\label{eq:intensity_rm}
{\cal P}_{rm}(\vec r,\tau)=p_m I_m(\vec r)\left|\hat e^*_r\hat e_m(\vec r)\right|^2{\cal G}^2_{rm}(\tau-\tau_{mr}),
\ee
The intensity distribution $I_m(\vec r)$ is obtained from this expression by
summation over a complete set of reference states (e.g. $|H\ra$ and $|V\ra$)
and then integrating over the temporal extent of the mode $\Lambda_m$, as
described earlier in this Chapter.

The polarization state of every mode, corresponding to a peak value of the
coherence function ${\cal G}(\tau-\tau_{mr})$ at $\tau=\tau_{mr}$, can be
reconstructed from the measurement of six different polarization states of the
reference beam $\hat e_r$. The number of measurements is reduced to four when
one accounts for the symmetry of the basis states.

It is convenient to introduce polarization distribution function
\be
{\cal U}_{rm}(\vec r)=\int_{\Lambda_m}d\tau\,{\cal P}_{rm}(\vec r,\tau)/\lp p_m I_m(\vec r)\rp=\left|\hat e^*_r\hat e_m(\vec r)\right|^2,
\ee
sensitive to the overlap of the reference polarization and spatially dependent
mode-polarization.

Straightforward calculations carried out for six spatially uniform
polarization states of the reference beam lead to the following set of
equations:
\bea 
{\cal U}_H&=&|E_x|^2,\quad {\cal U}_V=|E_y|^2,\nn
{\cal U}_+&=&\frac{1}{2}\lp |E_x|^2+|E_y|^2\rp+|E_xE_y|\cos\varphi,\nn
{\cal U}_-&=&\frac{1}{2}\lp |E_x|^2+|E_y|^2\rp-|E_xE_y|\cos\varphi,\nn 
{\cal U}_L&=&\frac{1}{2}\lp |E_x|^2+|E_y|^2\rp-|E_xE_y|\sin\varphi,\nn 
{\cal U}_R&=&\frac{1}{2}\lp |E_x|^2+|E_y|^2\rp+|E_xE_y|\sin\varphi, 
\eea
with the obvious symmetry identities ${\cal U}_0={\cal U}_H+{\cal U}_V={\cal U}_++{\cal U}_-={\cal U}_L+{\cal U}_R$.

The spatial distribution of the Stokes vector of the $m$-th mode is directly obtained from these relations, by averaging over the ensemble of measurements:
\bea
S_0(\vec r)&=&\la {\cal U}_H\ra+\la {\cal U}_V\ra=\la {\cal U}_0\ra,\nn
S_1(\vec r)&=&\la {\cal U}_H\ra-\la {\cal U}_V\ra=2\la {\cal U}_H\ra-\la {\cal U}_0\ra,\nn
S_2(\vec r)&=&\la {\cal U}_+\ra-\la {\cal U}_-\ra=2\la {\cal U}_+\ra-\la {\cal U}_0\ra,\nn
S_3(\vec r)&=&\la {\cal U}_R\ra-\la {\cal U}_L\ra=2\la {\cal U}_R\ra-\la {\cal U}_0\ra
\eea
In the case of a polarized light, the electric fields do not fluctuate within
the measurement ensemble, and the formalism simplifies due to the identity
$S_0^2=S_1^2+S_2^2+S_3^2$ limiting the polarization vector to the surface of
the Poincar\'e sphere.

This formalism has been used to obtain the spatial distribution of the Stokes
vector in the case of vortex states, as discussed in
Chapter~\ref{chapter:Experiments}.

\section{Phase-sensitive imaging}
\label{section:C2_Phase}

Spatial phase $\varphi_m(\vec r)$ is one of the key characteristics of
waveguide modes, illustrating the complex nature of electric field:
\be
\vec e_m(\vec r,\omega_0)=\sqrt{I_m(\vec r)}\exp\lb -i\varphi_m(\vec r)\rb
\hat e_m(\vec r),
\ee
Although in some trivial cases an irrelevant factor, the phase is essential
for understanding the vortex modes recently obtained in optical
waveguides.

In this section, I describe how one can use phase-sensitive C$^2$-imaging for
reconstruction of the spatial phase profile of waveguide
modes~\cite{Barankov2012_Phase}. In particular, the intensity distribution of
the interferometric signal contains overlap of the reference and $m$-th
waveguide modes:
\be\label{eq:intensity_phase}
I_{int}(\vec r,\tau)=2 \textbf{Re}\sum_m \alpha_m\vec e_{r}^*(\vec
r,\omega_0)\vec e_{m}(\vec r,\omega_0) G_{rm}\left(\tau-\tau_{mr}
\right)e^{-i\theta_{rm}},
\ee

The intensity distributions of the reference beam and the $m$-th mode, as well
as the polarization distributions can be obtained by the methods described
earlier in this Chapter. It becomes possible, therefore, as it follows from
Eq.~(\ref{eq:intensity_phase}), to introduce the phase function that
accounts only for the impact of the temporal dependence of the coherence
function and the spatially dependent phase of the $m$-mode:
\be\label{eq:cos_phase}
{\cal I}_{rm}(\vec r,\tau)={\cal G}_{rm}(\tau) \cos\lb\varphi_m(\vec
r)-\omega_0\tau+\psi_{rm}(\tau) +\theta_{rm}(\tau_{mr})\rb,
\ee
where, to simplify the notations, time variable was redefined:
$\tau-\tau_{mr}\to\tau$. Here, $\psi_{rm}(\tau)$ and ${\cal G}_{rm}(\tau)$ are
the phase and magnitude of the coherence function $G_{rm}(\tau)$, varying on
the time-scale significantly larger than the period $2\pi/\omega_0$ of the
electric field.

As it follows from Eq.~(\ref{eq:cos_phase}), one can use the following strategy to extract the spatial phase $\varphi_m(\vec r)$ from the phase function~(\ref{eq:cos_phase}). For a given position $\vec r_*$, we identify moments of time $\tau_{max}$ and $\tau_0$, such that $\tau_{max}-\tau_0\le\pi/(2\omega_0)$ and
\be
{\cal I}_{rm}(\vec r_*,\tau_{max})={\cal G}_{rm}(\tau_{max}),\quad {\cal I}_{rm}(\vec r_*,\tau_0)=0:\, \tau_{max}-\tau_0\le\pi/(2\omega_0).
\ee
As a result, one finds two complimentary functions
\bea
{\cal I}_{rm}(\vec r,\tau_{max})&=&{\cal G}_{rm}(\tau_{max})\cos\lb\varphi(\vec r)-\varphi(\vec r_*)\rb,\nn
{\cal I}_{rm}(\vec r,\tau_0)&=&{\cal G}_{rm}(\tau_0)\sin\lb\varphi(\vec r)-\varphi(\vec r_*)\rb,
\eea
which allow reconstruction of the spatial phase
\be
\varphi(\vec r)-\varphi(\vec r_*)=-i\,\ln\lb{\cal I}_{rm}(\vec r,\tau_{max})/{\cal G}_{rm}(\tau_{max})+i\,{\cal I}_{rm}(\vec r,\tau_0)/{\cal G}_{rm}(\tau_0)\rb.
\ee

This approach has been used for reconstruction of spatial phases of the vortex
modes, as demonstrated in Chapter~\ref{chapter:Experiments}.

\cleardoublepage

\chapter{C$^2$-imaging in experiments}
\label{chapter:Experiments}
\thispagestyle{myheadings}

In this Chapter, I discuss application of C$^2$-imaging for characterization
of several optical waveguides~\cite{Barankov2012_Review}. In
Section~\ref{section:TAP_LPG}, I demonstrate the consistency of the method in
retrieving modal content of a few-moded specialty fiber with an embedded
long-period Bragg grating, acting as a mode-converter. In this system, one
waveguide mode is converted into another as a function of wavelength, by the
resonant mode-coupling mechanism. Thus, an independent verification of the
modal content measured using C$^2$-imaging method becomes possible by an
independent spectroscopic measurement of the mode-conversion efficiency of the
mode-converter.

Next, in Section~\ref{section:Dispersion_compensation}, the
dispersion-compensation is discussed and applied for analysis of the modal
content of large-mode area fibers.

Polarization-sensitive modification of the method is applied in
Section~\ref{section:Polarization} for reconstruction of mode-specific
polarization distribution of vector modes propagating in a specialty fiber.
Specifically, the Stokes parameters of the modes are compared to the
theoretical predictions, providing the first to-date verification of the
polarization patterns of these modes in multimoded fibers.

The specialty fiber also supports two degenerate vector modes carrying orbital
angular momenta resulting in the characteristic discontinuity of their spatial
phase distribution. In Section~\ref{section:Phase}, I apply phase-sensitive
modification of C$^2$-imaging method for complete characterization of the
vortex modes, and obtain their relative weights, polarization and phase
properties.

In Section~\ref{section:LCF}, I employ C$^2$-imaging for identifying the
mechanism of resonant mode-coupling observed in coiled large-mode-area
leakage-channel fibers.

Regarding the experimental setup, the modal content of higher-order-mode fiber
with embedded long-period grating, and also the measurements of
large-mode-area fibers illustrating the concept of dispersion-compensation
were carried out using the basic experimental setup discussed in
Chapter~\ref{chapter:Formalism}. The polarization and phase-sensitive
measurements were carried out using a later-developed polarization-sensitive
modification of the setup discussed in Sections~\ref{section:Polarization}
and~\ref{section:Phase}.

The content of Sections~\ref{section:TAP_LPG}
and~\ref{section:Dispersion_compensation}, with some clarifying textual
changes, closely reproduces the corresponding parts of the journal
article~\cite{Schimpf2011}, to which I have been a key contributor. D. Schimpf
and I have contributed equally to work reported in these sections.

\section{Modal content of higher-order mode fiber with long-period grating}
\label{section:TAP_LPG}

The specialty higher-order mode (HOM) fiber supplemented by a
turn-around-point long-period grating (TAP-LPG) unit allows direct comparison
of modal content retrieved by C$^2$-imaging method to independent
spectroscopic measurements of the conversion efficiency. Since the
higher-order modes supported by this fiber demonstrate distinct dispersive
behavior, the same fiber is used to illustrate the impact of dispersion on the
interferometric signal.

\begin{figure}[!htbp]
\centering
\includegraphics[scale=.43]{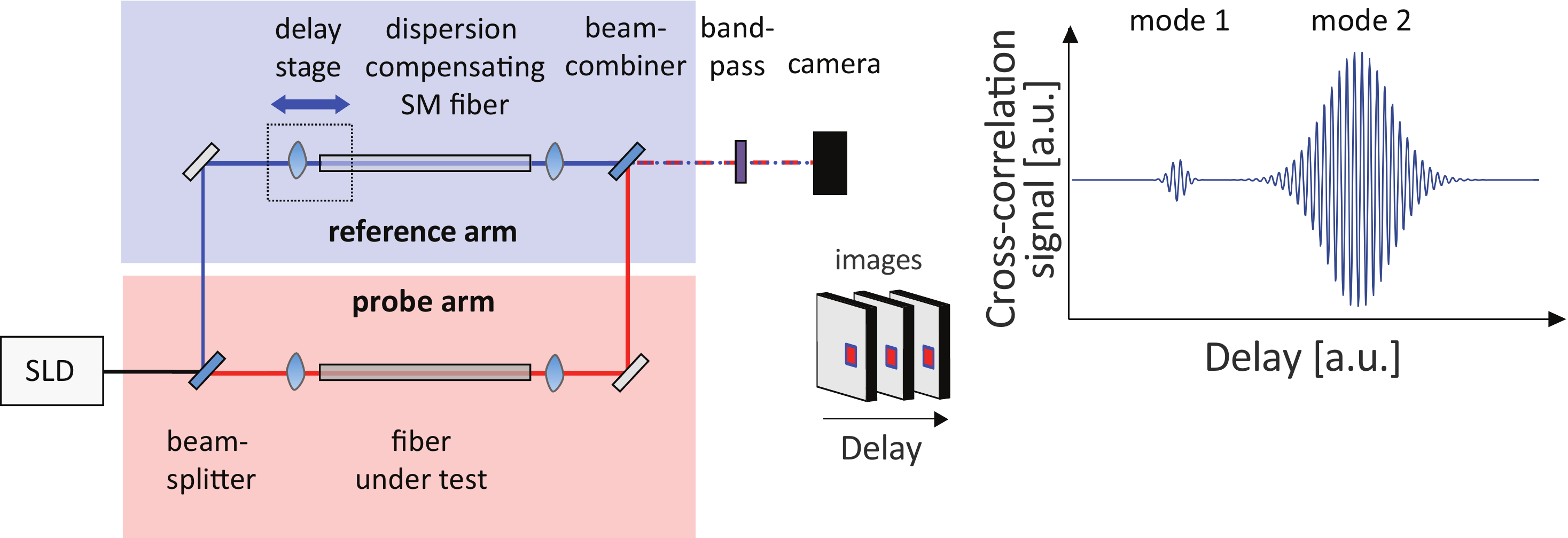}
\caption{Schematic of the experimental setup (SLD: superluminescent diode), and illustration of the cross-correlation trace expected at one pixel of the stack of images.}
\label{fig:setup}
\end{figure}

In the first set of experiments we employed free-space path for the reference
beam (the dispersion of optical elements positioned in this path was
negligible). A schematic of the setup is shown in Fig.~\ref{fig:setup}. The
few-mode fiber under test is the final element of a module (L=0.6 m)
consisting of a single-mode fiber, a TAP-LPG, and the higher-order mode fiber
(L=0.4 m)~\cite{Ramachandran2005,Jespersen2010}. In this fiber, the LPG
spectrum characterizes the mode conversion efficiency from the LP$_{01}$
core-mode to the LP$_{02}$ core-mode, which is used as an independent
reference for comparison against the ratio of modal weights obtained by
C$^2$-imaging method.

The source spectrum, provided by an LED, was filtered by a bandpass with the
central wavelength of about 780 nm and bandwidth of about 4 nm. Since the LPG
mode-conversion bandwidth exceeds 20 nm, the wavelength-dependent LPG spectrum
has a negligible impact on the spectra of HOM's.

\begin{figure}[!htbp]
\centering
\includegraphics[scale=.36]{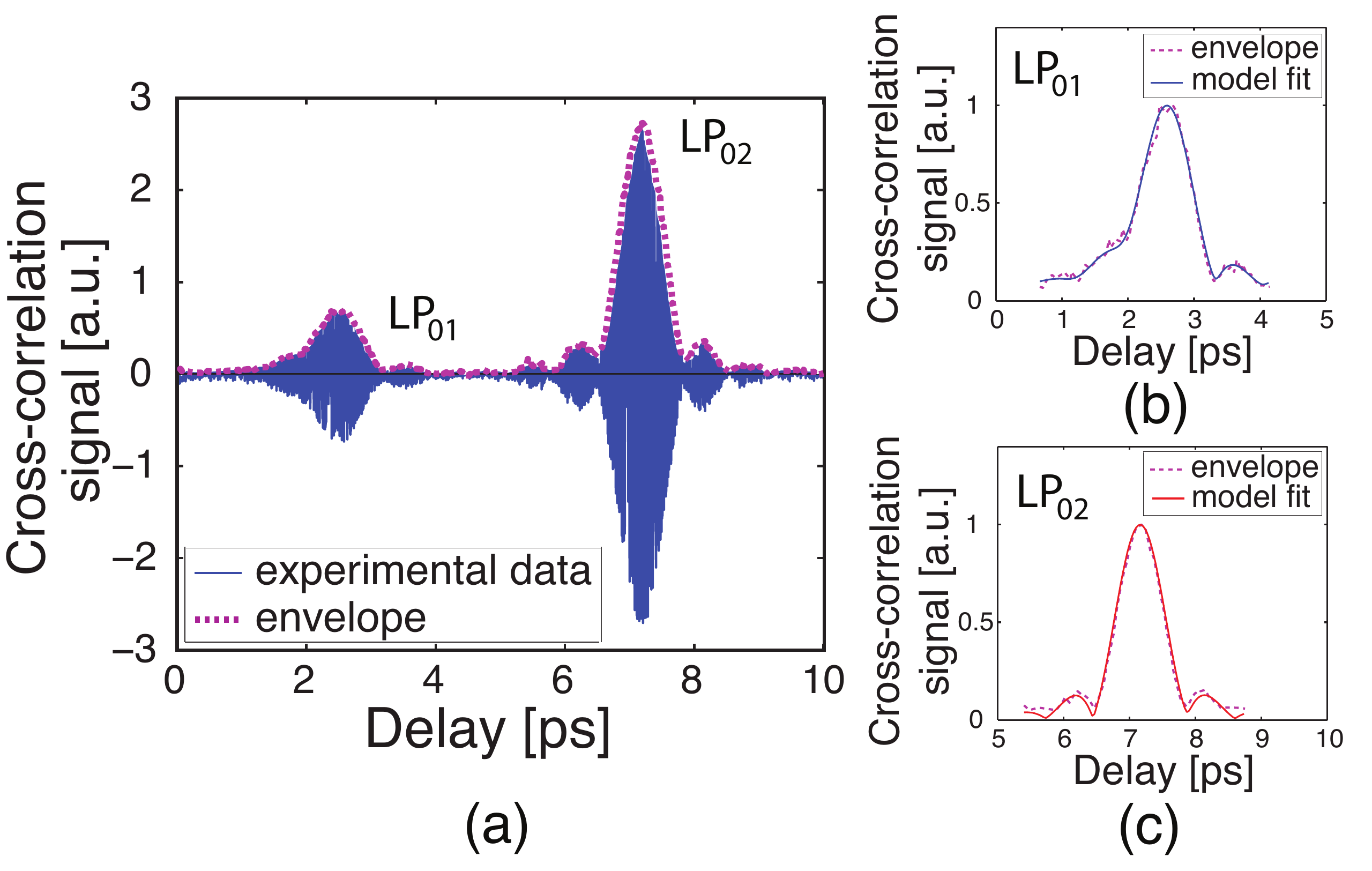}
\caption{(a): Cross-correlation trace for the entire image (data is offset corrected) for the bandpass centered at  $\lambda_{center}$=780 nm. (b) and (c): fit of the model to the envelope of the experimental data, for the first and second peak, corresponding to LP$_{01}$ and LP$_{02}$, respectively.}
\label{fig:trace}
\end{figure}

Figure~\ref{fig:trace}(a) shows an example of cross-correlation trace between
the reference field and the output of the fiber, integrated over all pixels of
the camera. The peaks in the trace correspond to the two different modes in
the HOM fiber. We analyze the data using elements of the theoretical analysis
discussed in Chapter~\ref{chapter:Formalism}.

The envelope of the cross-correlation trace is obtained from the stack of
images recorded by the camera, as shown in red color in Fig.~\ref{fig:trace}
(a). Then, the amplitude of the coherence function ${\cal G}_{rm}(\tau)$ is
fitted to the extracted envelope, separately for the two peaks, as
Figs.~\ref{fig:trace}(b) and (c) demonstrate.

The difference in shape of the two interference peaks reflects distinct modal
dispersion of the corresponding modes. The side-lobes around the dominant
peaks (clearly visible for the $LP_{02}$ mode) are due to the steep spectral
edges of the filtered spectrum. By fitting ${\cal G}_{rm}(\tau)$ to these
data, we obtain the relative group-delays and the dispersion of the two modes.

\begin{figure}[!t]
\centering
\includegraphics[scale=.40]{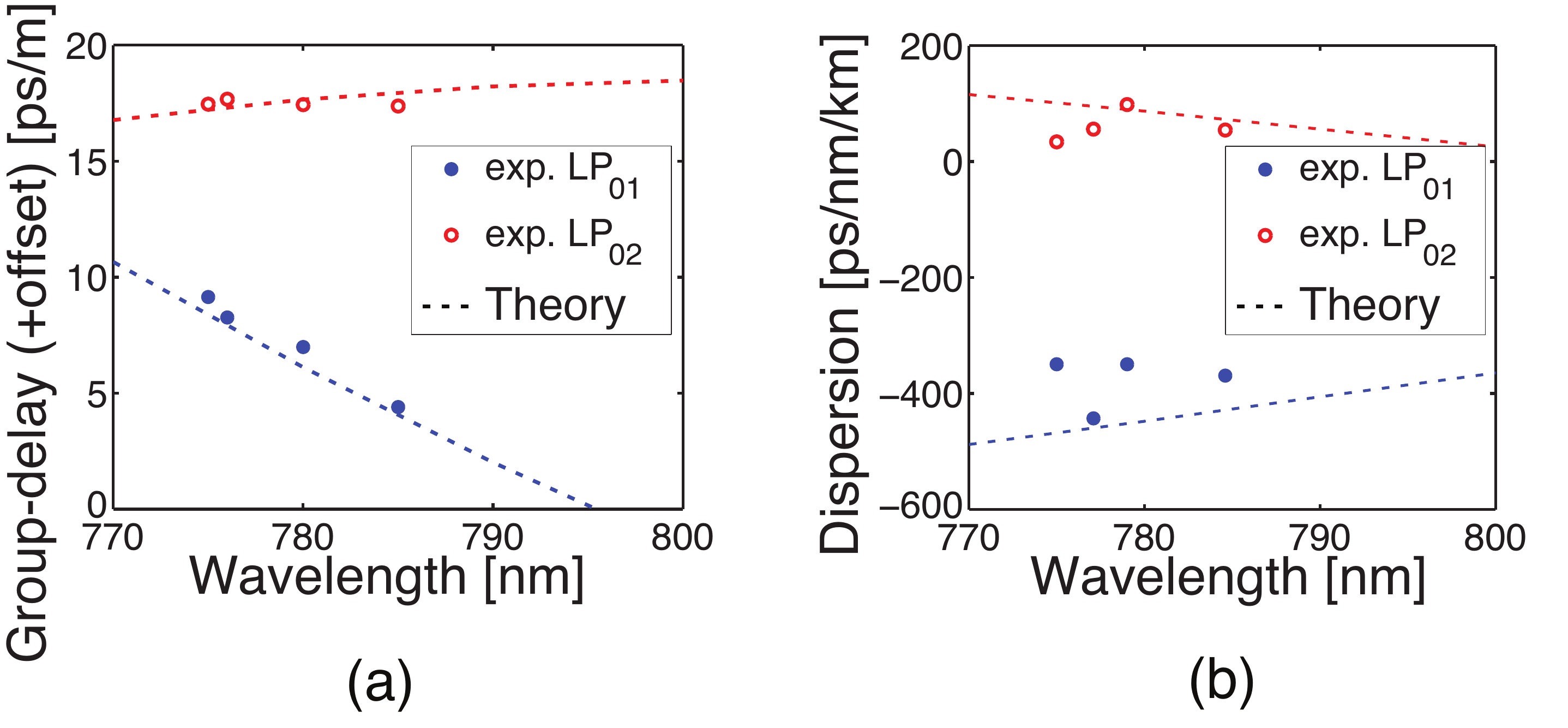}
\caption{(a) Group-delays, and (b) Dispersion values of the two modes as a function of center wavelength of the bandpass. }
\label{fig:GD}
\end{figure}

The dependence of the group-delay and dispersion on the wavelength is measured
by shifting the central wavelength of the filtered spectrum (achieved by
changing the incidence angle of light on the bandpass). The results are shown
in Figs.~\ref{fig:GD}(a) and (b). For comparison, we have also simulated the
modal properties of the tested fiber using a scalar mode-solver.

\begin{figure}[!t]
\centering
\includegraphics[scale=.53]{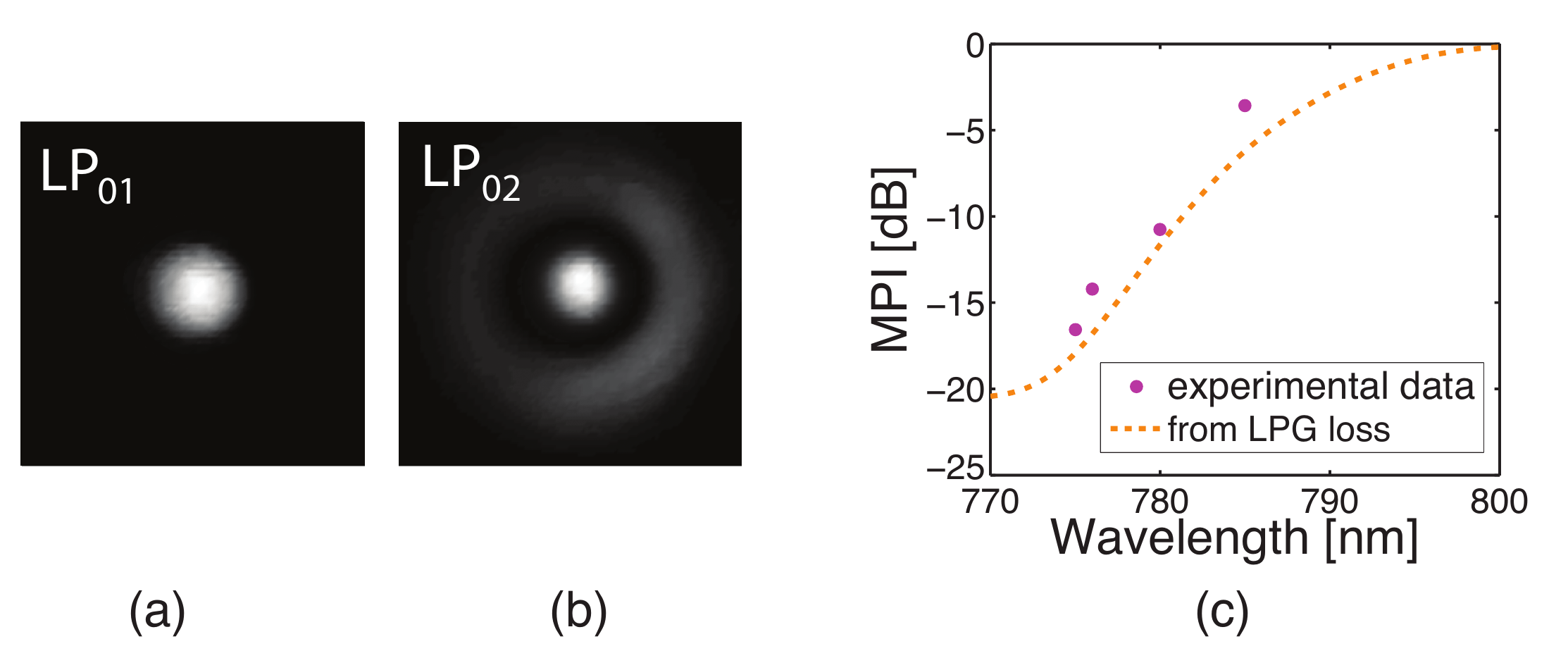}
\caption{(a) and (b), reconstructed $LP_{01}$ and $LP_{02}$-mode (gamma-adjusted) at a center wavelength of 780 nm, (c) multi-path interference (MPI) values as a function of center wavelength of the bandpass.}
\label{fig:MPI}
\end{figure}

Figures~\ref{fig:MPI}(a) and (b) show the reconstructed intensity
distributions of $LP_{01}$ and $LP_{02}$ modes that match the expected pattern
of the modes.

The relative (dispersion-corrected) weights of the normalized modes are
characterized using the multi-path interference (MPI) value
\be
MPI=10\log_{10}\lp p_m/p_{_{LP_{01}}}\rp,
\ee 
a useful measure of the relative power of a higher-order mode, $p_m$ to that
of the fundamental mode $p_{_{LP_{01}}}$.

We demonstrate the accuracy of the C$^2$ imaging in Fig.~\ref{fig:MPI}(c),
where the multi-path interference (MPI) value is plotted as a function of the
wavelength. The obtained MPI values match very well with the mode conversion
efficiencies independently measured by recording the LPG spectrum.

\section{Dispersion compensation for imaging of large-mode area fibers}
\label{section:Dispersion_compensation}

The advantages of dispersion compensation are illustrated in characterization
of large-mode-area (LMA) fibers supporting modes with similar dispersive
characteristics. As I argued in Chapter~\ref{chapter:Formalism}, in this case,
it is possible to achieve high temporal resolution of small intermodal group
delays, although, obviously, at the expense of spectral resolution.

The temporal resolution of C$^2$-imaging is related to the width of the
coherence function ${\cal G}_{mr}(\tau)$, which is determined by the spectral
properties of the light source. For a given shape of the spectrum, the
temporal resolution $\Delta \tau_{FWHM}$ (defined here as the FWHM of ${\cal
G}_{mr}(\tau)$) can be calculated as a function of the bandwidth of the
spectrum and the residual group-velocity dispersion (GVD) $\Delta
\varphi^{(2)}$.

\begin{figure}[!htb]
\centering
\includegraphics[scale=.4]{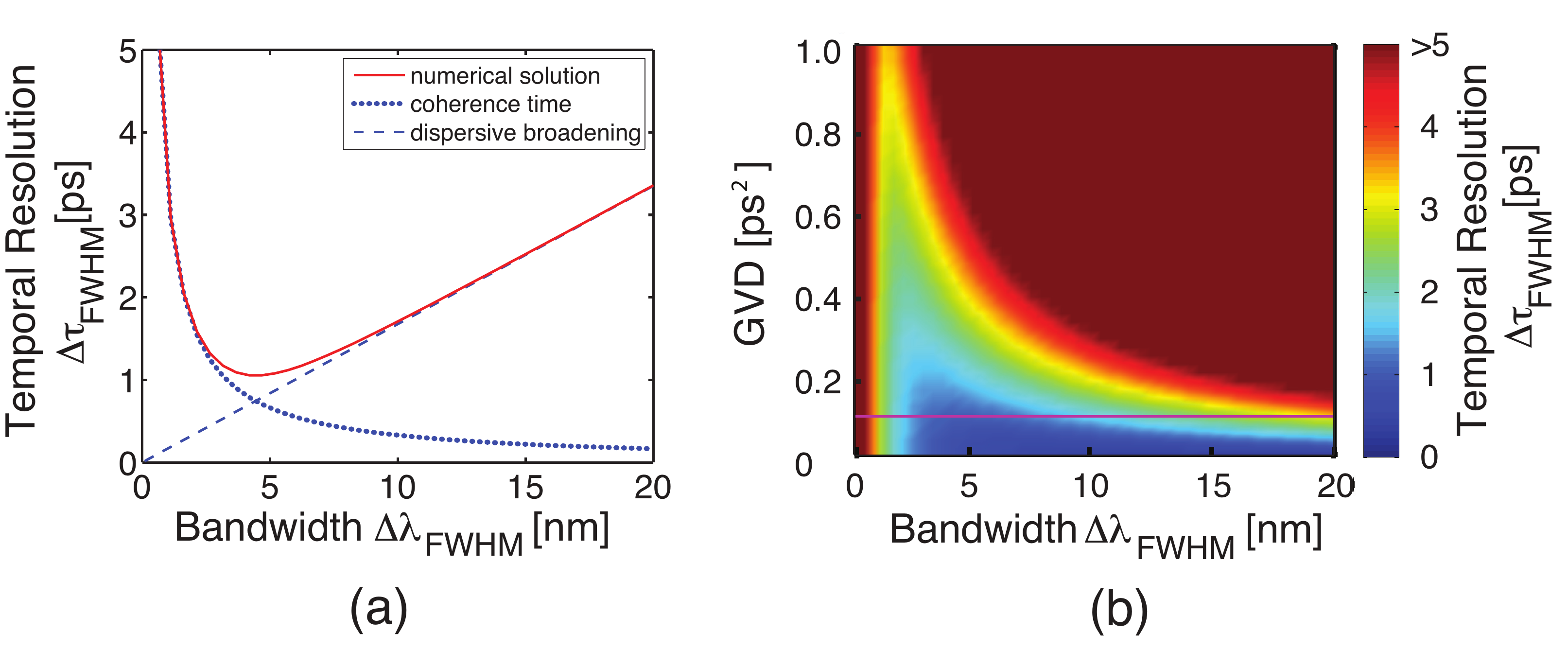}
\caption{(a) Temporal resolution as a function of the FWHM spectral bandwidth
of a Gaussian spectrum (for GVD value of $\varphi^{(2)}=0.1 ps^2$), (b) and as
a function of both FWHM bandwidth and GVD.}
\label{fig:resolution}
\end{figure}

We modeled the dependence of the temporal resolution for a Gaussian spectrum
centered at 1060 nm and residual GVD of $\Delta \varphi^{(2)}=0.1 ps^2$
(corresponding to about 4 m of LMA fiber and in the absence of the dispersion
compensating fiber in the reference path). The results are shown in
Fig.~\ref{fig:resolution}(a).

For small bandwidths, dispersion plays a minor role, and the resolution is
governed by the coherence time, which can be defined as $\Delta
\tau_{FWHM}^{coh} = 8\ln(2)/\Delta\omega_{FWHM}$, where $\Delta\omega_{FWHM}$
is the width of the spectrum at the angular frequency ($
\Delta\omega_{FWHM}\approx 2\pi c_0\Delta\lambda_{FWHM}/\lambda_0^2$).

In the absence of dispersion, broad spectra may be used to achieve maximal
temporal resolution. However, when the interferometer is not dispersion
balanced, the broadening of cross-correlation signal is dominated by the
dispersion, i.e. $\Delta \tau_{FWHM}^{disp} = \Delta \varphi^{(2)}
\Delta\omega_{FWHM}$. Nonetheless, even in this case, it is possible to
identify an optimal bandwidth of the spectrum (i.e. appropriate choice of the
bandpass) corresponding to the optimal resolution, as
Fig.~\ref{fig:resolution}(b) clearly illustrates. The horizontal line in Fig.
\ref{fig:resolution}(b) highlights the parameter configuration shown in Fig.
\ref{fig:resolution}(a).

The temporal resolution may be noticeably improved in a dispersion-matched
interferometer, a well-known result in optical coherence tomography
(OCT)~\cite{Wojtkowski2004}. In our setup, we realize this idea by employing a
single-mode fiber in the reference path, which balances the dispersion of the
test fiber. Certainly, the dispersion can be exactly matched for a single mode
or group of modes with similar dispersive characteristics. For the other
modes, the residual dispersive phases cause broadening of the corresponding
coherence peaks in the cross-correlation trace.

The shape and smoothness of the spectrum also have an impact on the
cross-correlation trace, and thus, affect the temporal resolution. These
effects have also been discussed in the context of OCT~\cite{Boer}.

For characterization of large-mode area (LMA) fibers, in which all the modes
have similar magnitudes of chromatic dispersion and the relative delays are on
a picosecond (or less) timescales, the results of dispersion-matching are
particularly illuminating, as we demonstrate below.

We demonstrate dispersion compensation for characterization of the
polarization-maintaining LMA fiber with a core diameter of approximately 27.5
$\mu m$ and a numerical aperture (NA) of 0.062, having the length of 5 m. We
use a polarizer and a pair of half-wave plates to ensure launch of the beam
into one of the dominant polarization states of the fiber. The dispersion is
matched by using 4.08 meters of single-mode HI-1060 fiber in the reference
path. This length has been determined by cutting the input of the reference
fiber until the width of the dominant peak in the cross-correlation trace
matched the coherence function calculated for the full spectrum of the source.

\begin{figure}[!htb]
\centering
\includegraphics[scale=.43]{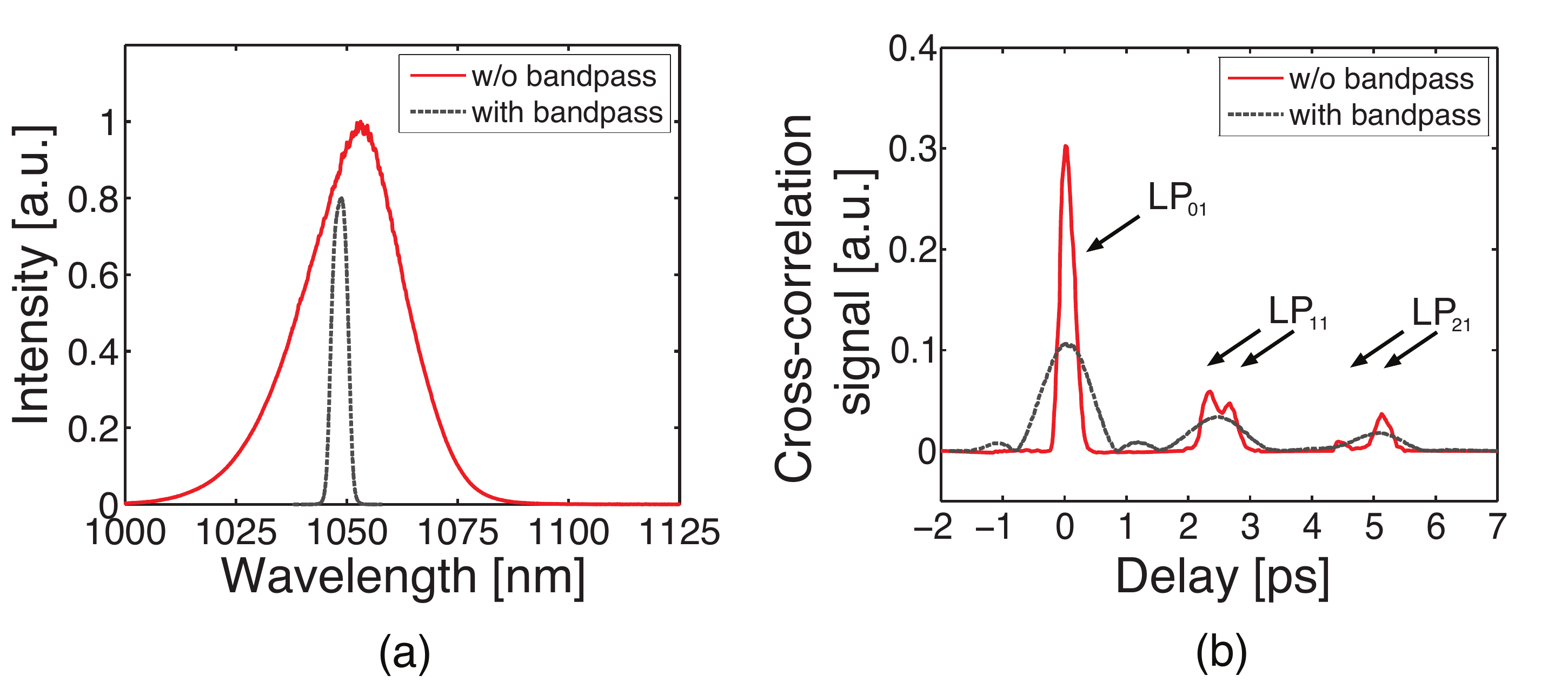}
\caption{(a): Full spectrum of the source and the 5-nm bandpass, (b): the envelopes of the cross-correlation traces.}
\label{fig:xtrace}
\end{figure}

\begin{figure}[!htb]
\centering
\includegraphics[scale=.25]{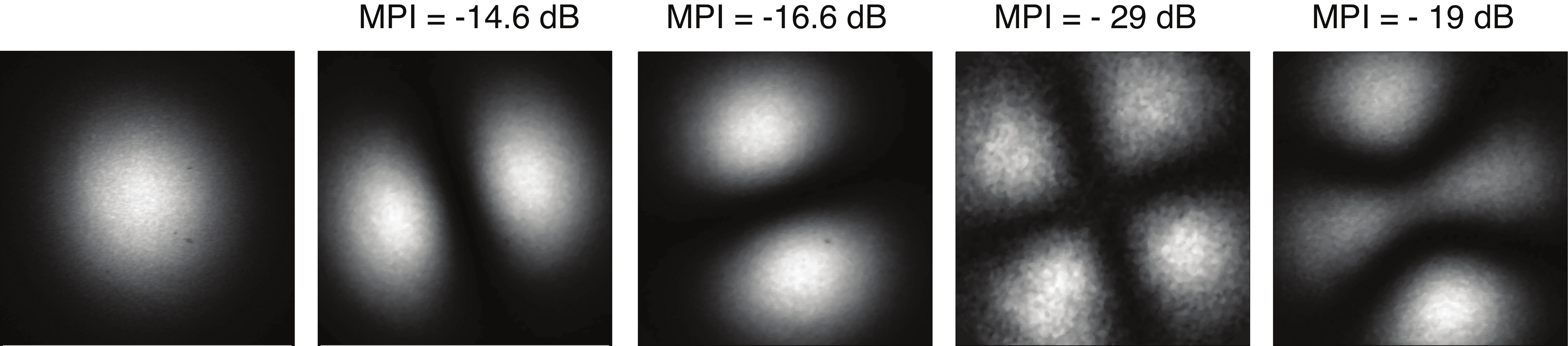}
\caption{Reconstructed mode profiles in the order of temporal delays shown in the cross-correlation trace of Fig.~\ref{fig:xtrace} (b) for the case of the full spectrum of the source.}
\label{fig:allmodes}
\end{figure}

\begin{figure}[!htb]
\centering
\includegraphics[scale=.42]{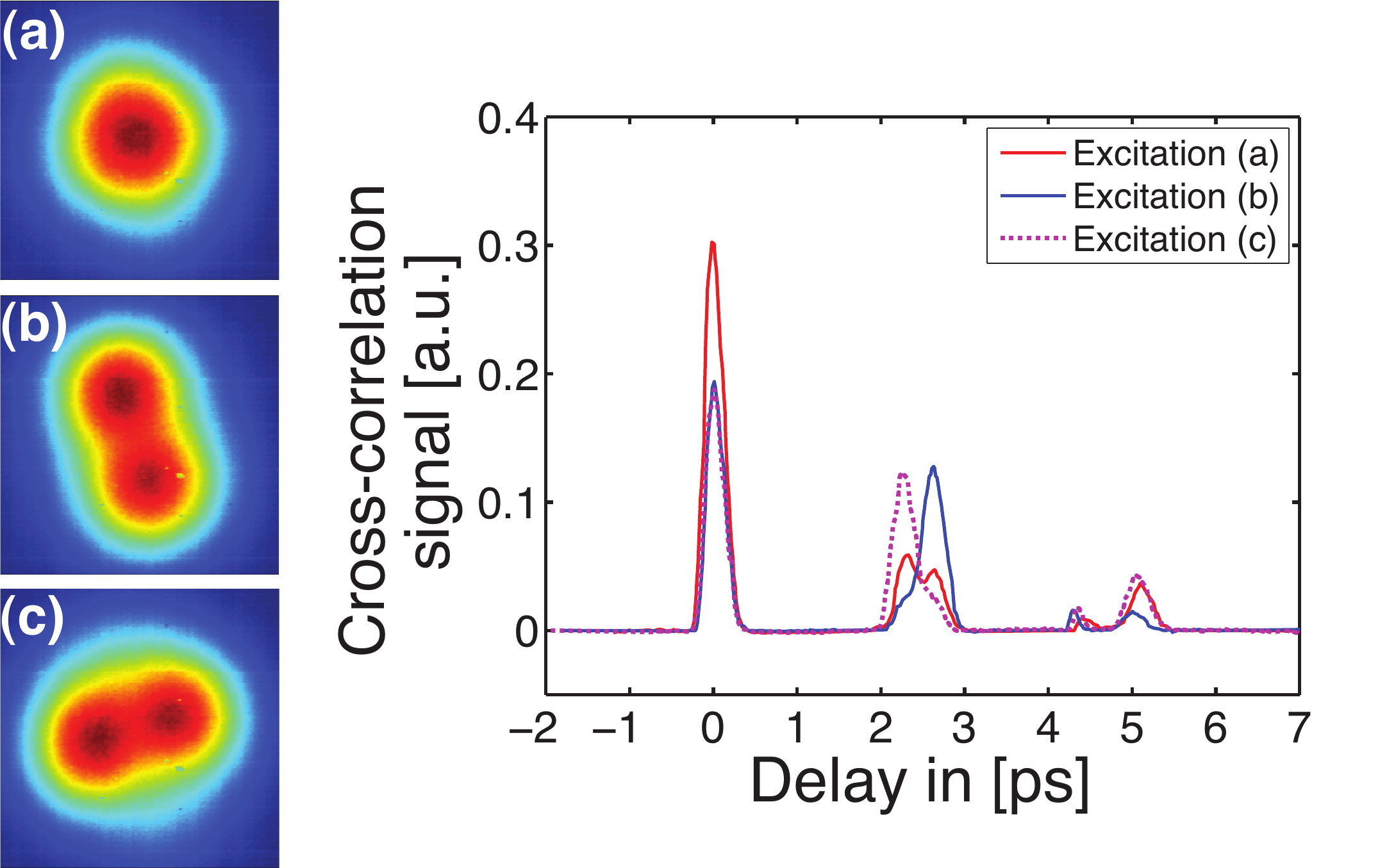}
\caption{(a-c) Output near-field images of the tested fiber, obtained for different in-coupling conditions, and corresponding changes in the cross-correlation trace.}
\label{fig:modes_ex}
\end{figure}

The effect of dispersion-matching is clearly visible when comparing the
cross-correlation traces, obtained using the full spectrum of the source in
Fig.~\ref{fig:xtrace}(a), and those recorded using a spectrum filtered by a
5-nm bandpass filter in Fig.~\ref{fig:xtrace}(b). Most importantly, dispersion
compensation combined with the use of full spectrum reveals the pairs of
$LP_{11}$ and $LP_{21}$ modes, which overlap in the traces obtained using the
filtered spectrum. The corresponding temporal resolution approaches values
smaller than 300 fs. The impact of the shape of the spectrum on the
cross-correlation trace is also noticeable: since the filtered spectrum has
steep edges, the corresponding cross-correlation trace shows characteristic
ringing (especially around the peak of the $LP_{01}$-mode). In contrast, the
smooth full spectrum results in a cross-correlation trace without spectral
artifacts.

Fig.~\ref{fig:allmodes} demonstrates intensity distributions of all modes,
arranged in the order of their time delay in Fig.~\ref{fig:xtrace}(b). The
dispersion-corrected MPI values of the modes shown in Fig.~\ref{fig:xtrace}(b)
(full spectrum) are estimated as -14.6 dB and -16.6 dB for the two $LP_{11}$
modes, and -29 dB and -19 dB for the two $LP_{21}$ modes. Interestingly, the
dimensionless V-parameter of the test fiber is about 5, so the next higher
$LP_{02}$-mode should also be present in the trace, but it is not observed in
our measurements. Coiling of the fiber to a diameter of around 30 cm may be
responsible for stripping off of the mode.

It is instructive to study the change of the modal pattern for different
in-coupling conditions. A qualitative observation based upon the image of the
output of the fiber in Fig.~\ref{fig:modes_ex}(a), obtained when the reference
beam is blocked, suggests fundamental-mode operation. Conversely,
C$^2$-imaging reveals that in fact the beam contains a significant amount of
power in the higher-order modes. The evolution of the modal content for other
in-coupling conditions measured by C$^2$-imaging is illustrated in
Fig.~\ref{fig:modes_ex}. Here, the near-field image of the fiber-output in
Fig.~\ref{fig:modes_ex}(b) indicates a stronger excitation of the ``slow"
$LP_{11}$-mode, while Fig.~\ref{fig:modes_ex}(c) shows a stronger ``fast"
$LP_{11}$-mode. Interestingly, in these two situations, $LP_{01}$-mode still
carries most of the modal power.

These measurements also reveal that temporal splitting between the $LP_{21}$
modes is more pronounced compared to the splitting between the $LP_{11}$
modes. It is an indication that modes with higher orbital angular momentum
(i.e. $LP_{lm}$ modes with higher $l$) are more susceptible to the
birefringence effects in this polarization maintaining
fiber~\cite{Golowich2005}.

\section{Polarization reconstruction of vector modes}
\label{section:Polarization}

One of the major advantages of C$^2$-imaging method compared to other
waveguide characterization techniques~\cite{Nicholson2009} is the flexibility
in choosing the polarization states of reference and signal beams
independently of one another, as described in Chapter~\ref{chapter:Formalism}.

In this section, I employ this advantage of the method to characterize
polarization properties of vector modes $TM_{01}$, $HE_{21}$, and
$TE_{01}$~\cite{Barankov2012_Polarization} propagating in the specialty fiber,
where the degeneracy between the modes is lifted by the ingenious design of
the index profile. The analysis is based on the formalism developed in
Chapter~\ref{chapter:Formalism}.

The four vector modes propagating in the fiber can be represented in the following way:
\bea\label{eq:vector_modes}
\vec e_{TM_{01}}&=&f(r)\lp\hat x\cos\varphi+\hat y\sin\varphi\rp,\nn
\vec e_{TE_{01}}&=&f(r)\lp\hat x\sin\varphi-\hat y\sin\varphi\rp,\nn
\vec e_{+}&=&f(r)e^{i\varphi}\lp\hat x+i\hat y\rp/\sqrt{2},\nn
\vec e_{-}&=&f(r)e^{-i\varphi}\lp\hat x-i\hat y\rp/\sqrt{2},
\eea
where $f(r)$ is the radial distribution of the modes that has characteristic
``donut'' shape; $\hat x$ and $\hat y$ are the unit vectors in the image
plane. It is convenient to employ the cylindrical coordinate system,
$x=r\cos\varphi$, $y=r\sin\varphi$, where $r$ is measured from the axis of the
fiber, and $\varphi$ is the azimuthal angle.

The last two modes $\vec e_{\pm}\sim e^{\pm\varphi}$ are the degenerate
orbital-angular-momentum (OAM) states characterized by the spatial phase
$\varphi$ that encodes the orbital momentum $L_z=\pm 1$. Due to the spectral
degeneracy, an arbitrary linear superposition of the modes defines $HE_{21}$
mode:
\be\label{eq:HE_21}
\vec e_{HE_{21}}=c_+\vec e_{+}+c_-\vec e_{-},
\ee
where $c_+$ and $c_-$ are the complex-valued amplitudes of the modes normalized to unity, $|c_+|^2+|c_-|^2=1$. The amplitudes depend on the in-coupling conditions to the fiber.

\begin{figure}[htb]
 	\includegraphics[width=14cm]{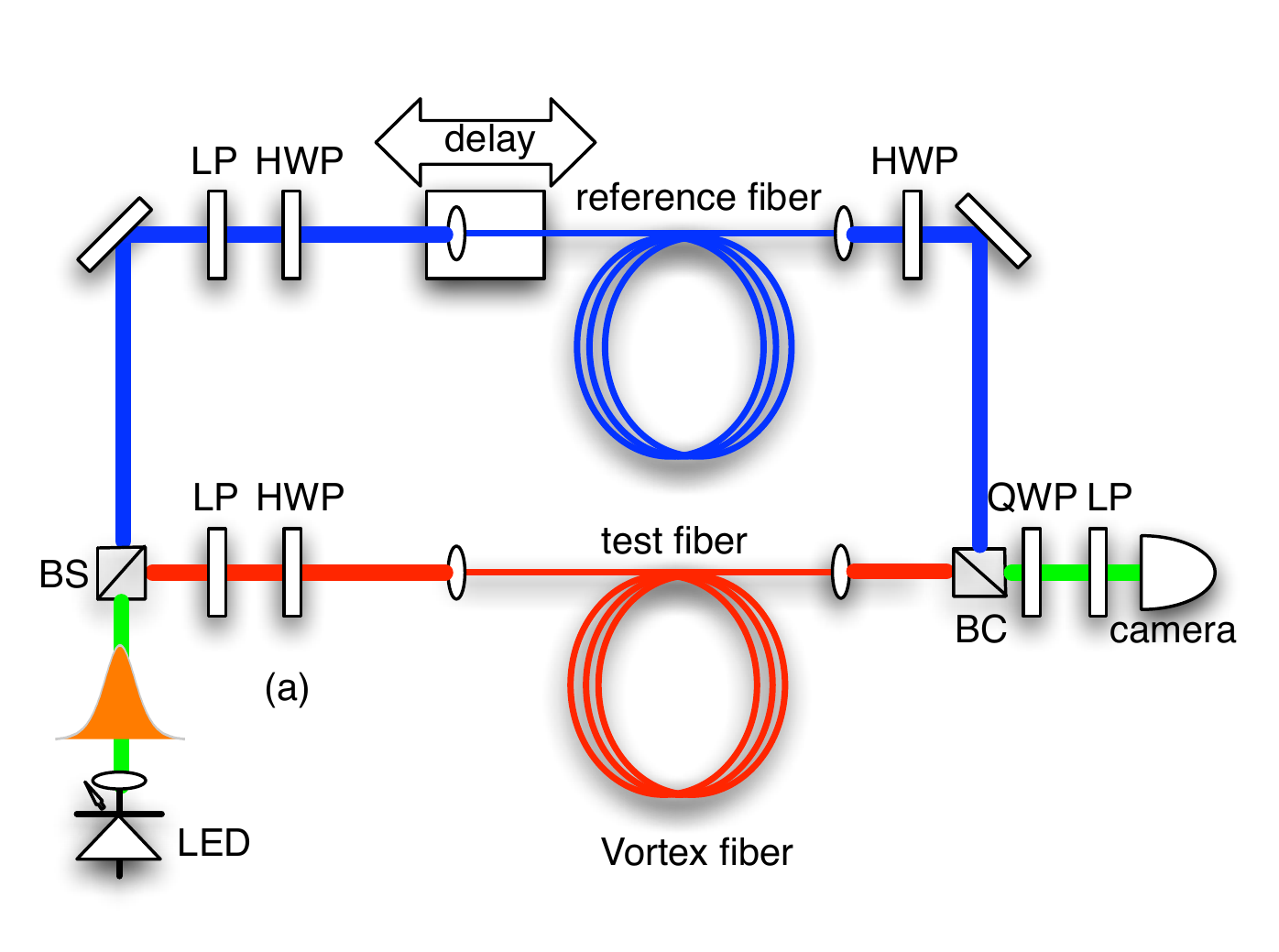}
  \caption{C$^2$-imaging setup: LED – light-emitting diode, LP-linear polarizer, HWP – half-wave plate, QWP – quarter-wave plate, BS – beam-splitter, BC – beam combiner.}
  \label{fig:setup_polarization}
\end{figure}

The polarization-sensitive C$^2$-imaging method is implemented using the
experimental setup shown in Fig.~\ref{fig:setup_polarization}. I employ a
pair of linear polarizers and half-wave plates to control the polarization
state of the reference beam. The in-coupling conditions to the test fiber is
also determined by a similar set of polarizing elements. Polarization
selection at the output of the system is provided by a combination of a
quarter-wave plate and a linear polarizer.

I used an LED source centered at about $\lambda=1524\,nm$ with spectral
width of $\sim 30\,nm$. The output beam of the test fiber focused at the
camera interferes with the collimated reference beam having well-defined
linear polarization. The in-coupling conditions remain the same throughout our
measurements. The length of the polarization-maintaining reference fiber is
chosen to match the optical paths of the two beams and also to compensate for
the modal dispersion of the test fiber. The effect of dispersion has been also
minimized by using a spectral filter with a central wavelength of
$\lambda\approx 1549\,nm$ and width $\lambda_{FWHM}\approx 6\,nm$. The camera
detects a cross-correlation signal as a function of the group delay monitored
by a computer-controlled delay stage. In the measurements, the
cross-correlation trace has been recorded for three pairs of orthogonal
polarization states of the reference beam: linear vertical, linear horizontal,
linear at $+\pi/4$ radians, linear at $-\pi/4$ radians, left-circular and
right-circular polarizations. The envelopes of the cross-correlation traces
have been analyzed to study the polarization properties and the relative power
of the modes.

\begin{figure}[!htb]
\centering
\includegraphics[scale=.7]{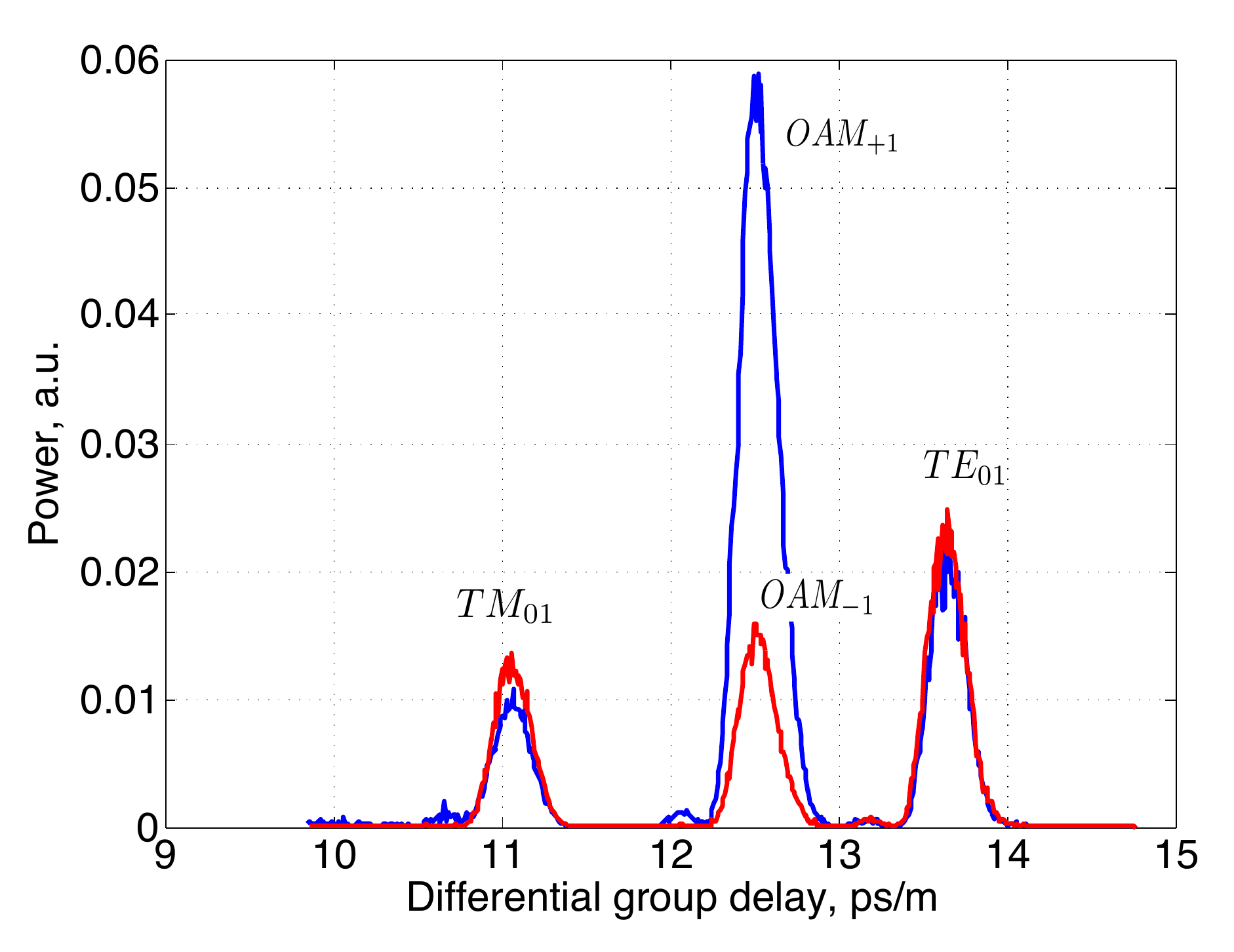}
\caption{Relative modal power as a function of group delay: blue line – right-circular polarization, red line – left- circular polarization}
\label{fig:circular_trace}
\end{figure}

Among the six polarization states used in the experiment, the left-circular
and right-circular states are particularly important for identifying the
topological charge of the OAM state. Specifically, as
Fig.~\ref{fig:circular_trace} illustrates, $TE_{01}$ and $TM_{01}$ modes are
insensitive to changes in orientation of the circular polarizer, as dictated
by their spatial and polarization properties. In contrast, the two OAM states
with the orbital angular momentum $L_z=\pm 1$, have the right and left
circular polarizations, correspondingly. Therefore, one can detect each of
these states using a circular polarizer, as shown in
Fig.~\ref{fig:circular_trace}. The peak values of the thus-recorded
cross-correlation traces contain the relative weights of the two OAM’s, and
also the weights of the other two modes.

\begin{figure}[!htb]
\centering
\includegraphics[scale=.7]{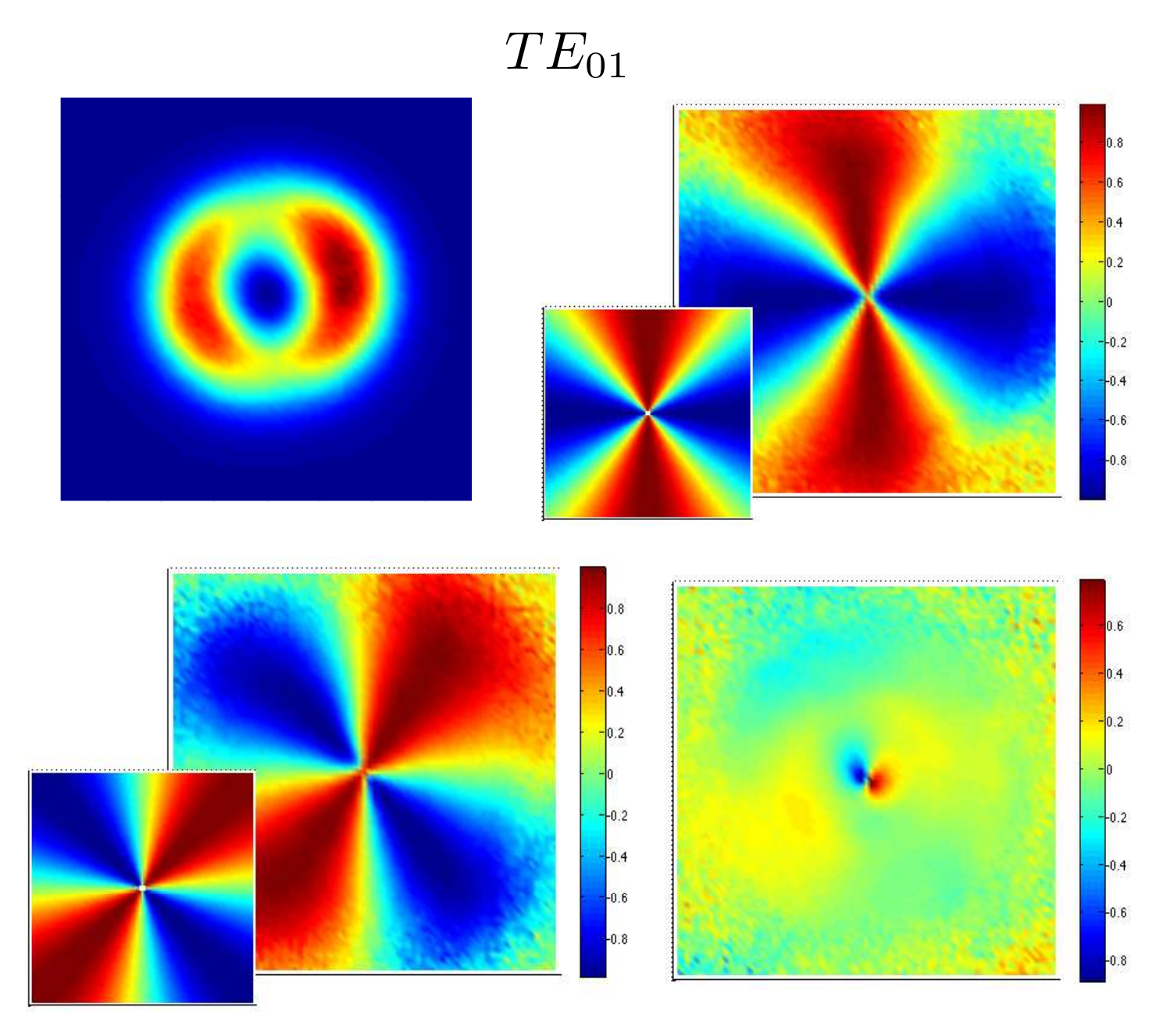}
\caption{Reconstruction of Stokes parameters of $TE_{01}$ mode:  (a) component $S_0$, (b) component $S_1$, (c) component $S_2$, (d) component $S_3$. {\it Inset:} theoretical prediction.}
\label{fig:stokes_TE01}
\end{figure}

\begin{figure}[!htb]
\centering
\includegraphics[scale=.7]{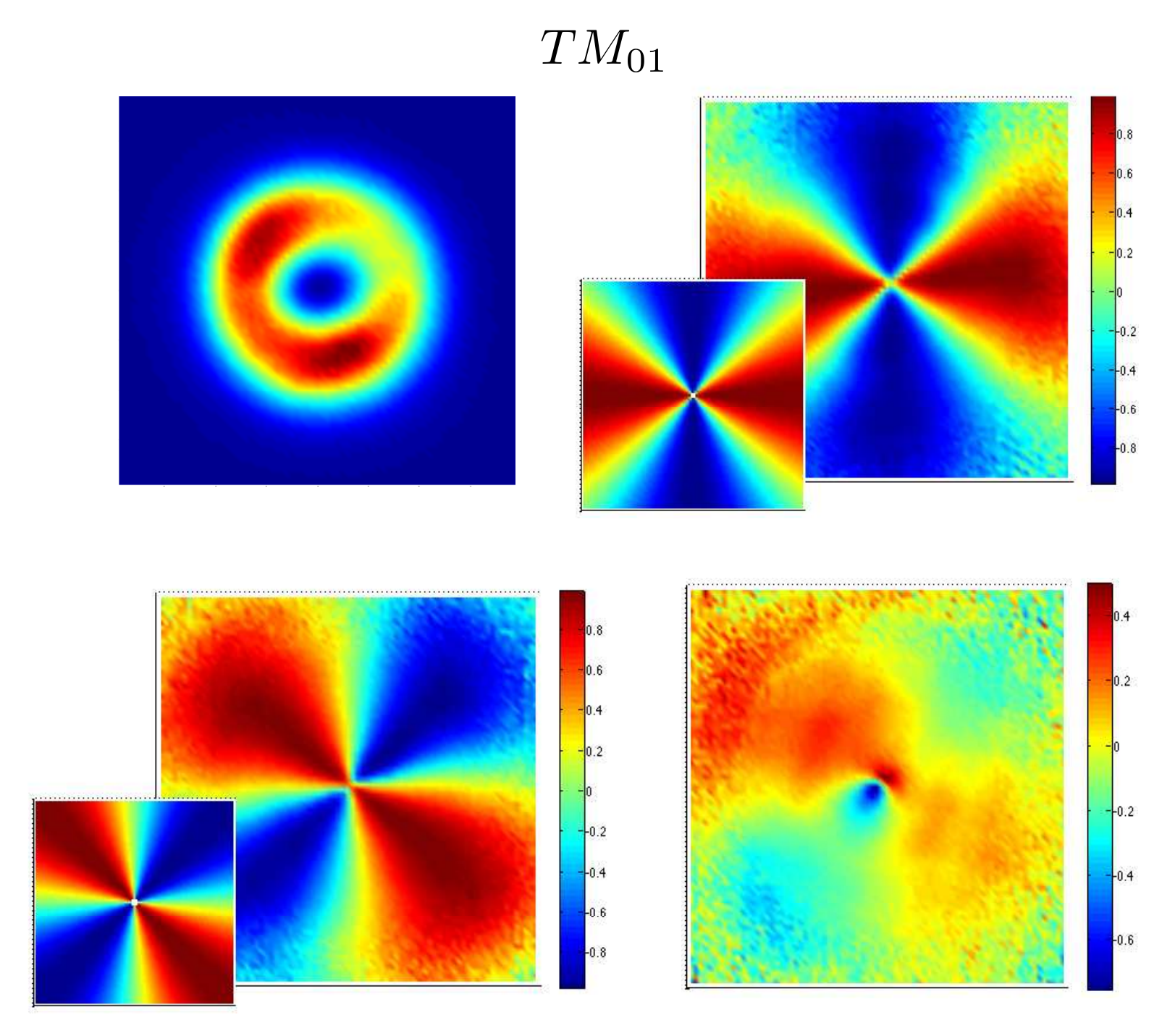}
\caption{Reconstruction of Stokes parameters of $TM_{01}$ mode:  (a) component $S_0$, (b) component $S_1$, (c) component $S_2$, (d) component $S_3$. {\it Inset:} theoretical prediction.}
\label{fig:stokes_TM01}
\end{figure}

Polarization-sensitive imaging method described in
Chapter~\ref{chapter:Formalism} allows complete reconstruction of the Stokes
parameters of $TM_{01}$, $TE_{01}$, and $HE_{21}$ modes and direct comparison
to the theoretical predictions based upon polarization patterns shown in
Eqs.~(\ref{eq:vector_modes}) and~(\ref{eq:HE_21}).

Specifically, for $TM_{01}$ the Stokes vector components are
\be
S_0(\vec r)=f(r),\,S_1(\vec r)=-f(r)\cos (2\varphi),\, 
S_2(\vec r)=f(r)\sin (2\varphi),\,S_3(\vec r)=0.
\ee
Similarly, for $TE_{01}$, one obtains
\be
S_0(\vec r)=f(r),\,S_1(\vec r)=f(r)\cos (2\varphi),\, 
S_2(\vec r)=-f(r)\sin (2\varphi),\,S_3(\vec r)=0.
\ee
In the most interesting case of $HE_{21}$ mode, the Stokes parameters $S_1$ and $S_2$ encode the interference term $c_+c^*_-=|c_+c_-| e^{i\alpha}$ of the two degenerate OAM states
\bea
S_0(\vec r)&=&f(r),\,S_1(\vec r)=2|c_+c_-|f(r)\cos (2\varphi+\alpha),\nn
S_2(\vec r)&=&-2|c_+c_-|f(r)\sin (2\varphi+\alpha),\,S_3(\vec r)=|c_+|^2-|c_-|^2
\eea

\begin{figure}[!htb]
\centering
\includegraphics[scale=.7]{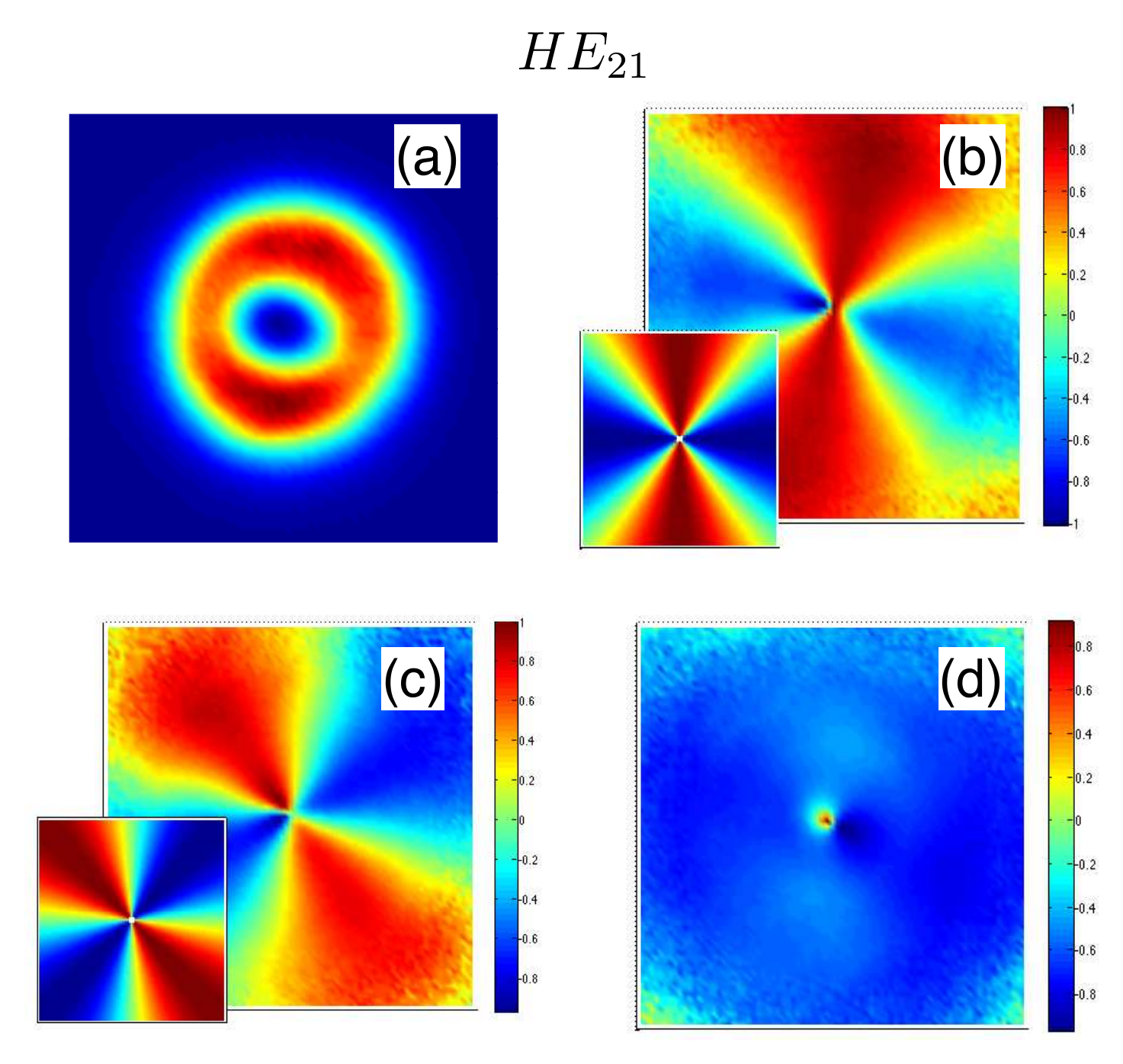}
\caption{Reconstruction of Stokes parameters of $HE_{21}$ mode: (a) component $S_0$, (b) component $S_1$, (c) component $S_2$, (d) component $S_3$. {\it Inset:} theoretical prediction}
\label{fig:stokes_HE21}
\end{figure}

The intensity distribution of each mode, given
by the component $S_0$ of the spatially dependent Stokes vector, as expected, demonstrates characteristic donut shape, as illustrated in Figs.~\ref{fig:stokes_TE01},~\ref{fig:stokes_TM01}, and~\ref{fig:stokes_HE21}. 

In summary, the reconstructed Stokes parameters for all vector modes compare
very well with the theoretical predictions. In addition, phase patterns $S_1$
and $S_2$ in Fig.~\ref{fig:stokes_HE21} reveal the relative phase of the
degenerate OAM states $\vec e_\pm$.

\section{Spatial phase reconstruction of vortex modes}
\label{section:Phase}

Polarization properties of OAM states shown in Eqs.(\ref{eq:vector_modes})
suggest a simple method to measure their modal power. Indeed, by projecting
the output beam to the left and right circular polarized states, one obtains
the cross-correlation traces of the states independently of one another.
Moreover, since the other two vector states, $TM_{01}$ and $TE_{01}$, are
insensitive to the change of the circular polarization, these states can be
easily identified and separated in the cross-correlation traces, as shown in
Fig.~\ref{fig:circular_trace}.

One of the key characteristics of OAM states
\bea\label{eq:oam_modes}
\vec e_{+}&=&f(r)e^{i\varphi}\lp\hat x+i\hat y\rp/\sqrt{2},\nn
\vec e_{-}&=&f(r)e^{-i\varphi}\lp\hat x-i\hat y\rp/\sqrt{2},
\eea
is the phase factor $e^{\pm i\varphi}$ corresponding to the two eigen-values of orbital momentum $L_z=\pm 1$ along the direction of propagation.

In this section, I apply the phase-sensitive C$^2$-imaging to measure the
phase distribution of the vortex states~\cite{Barankov2012_Phase}. First, I
project the interferometric signal onto the right- and left- circular
polarization states to select the two vortex states independent of one
another. The resulting interferometric trace contains the information about
the spatial phase that can be retrieved using simple phase-stepping algorithm
developed in Chapter~\ref{chapter:Formalism}.

The cross-correlation signal is measured with high temporal resolution
corresponding to a fraction of the wavelength in spatial translation of the
delay stage. The phase-stepping algorithm is capable of retrieving the phase
information even for relatively noisy conditions.

\begin{figure}[!htb]
\centering
\includegraphics[width=10cm]{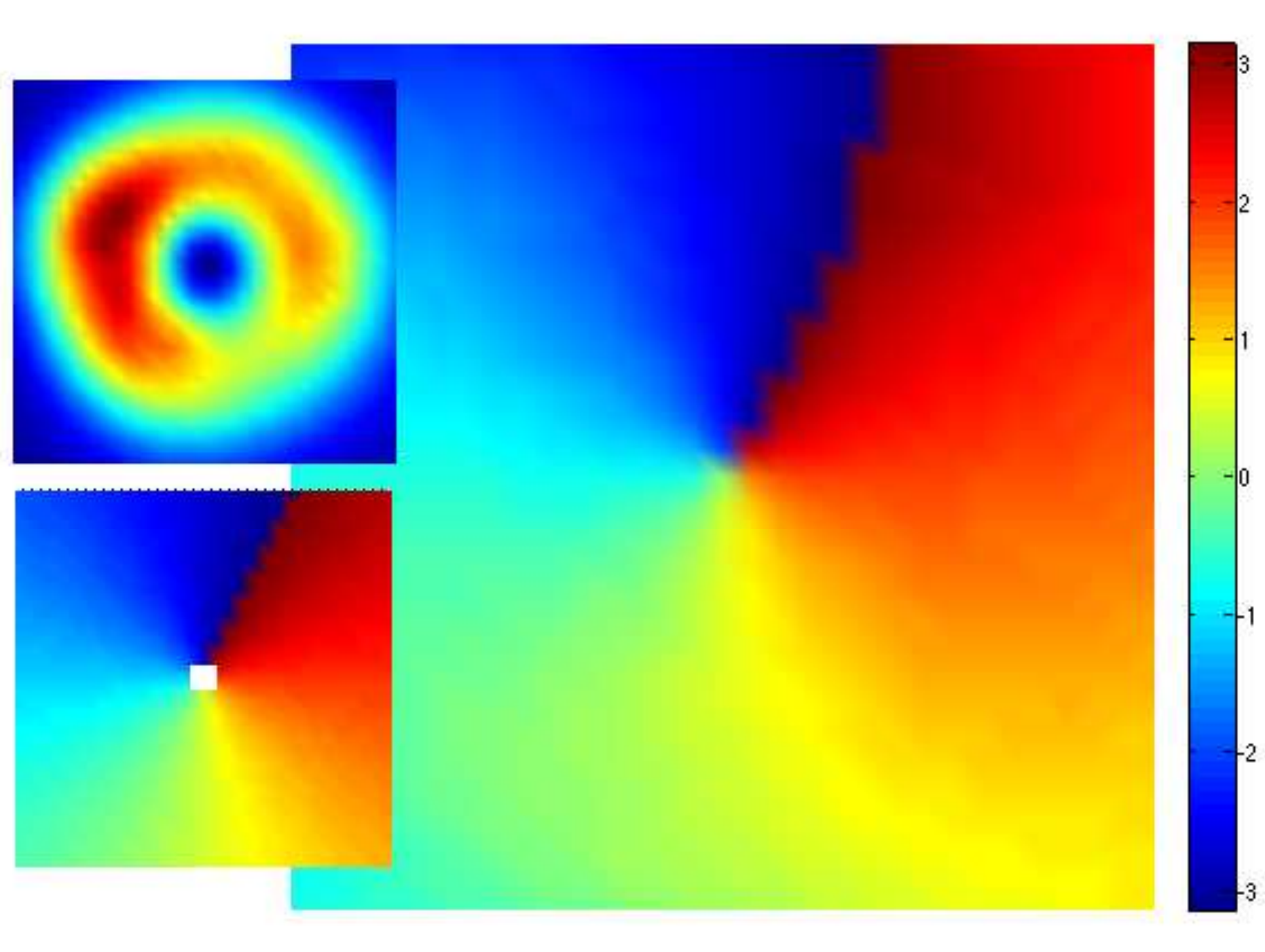}
\caption{Spatial phase of OAM state with $L_z=+1$ that shows characteristic phase wrapping in the counter-clockwise direction. {\it Insets:} (a) Intensity distribution of the mode, (b) theoretical distribution of spatial phase}
\label{fig:oam_plus}
\end{figure}

\begin{figure}[!htb]
\centering
\includegraphics[width=10cm]{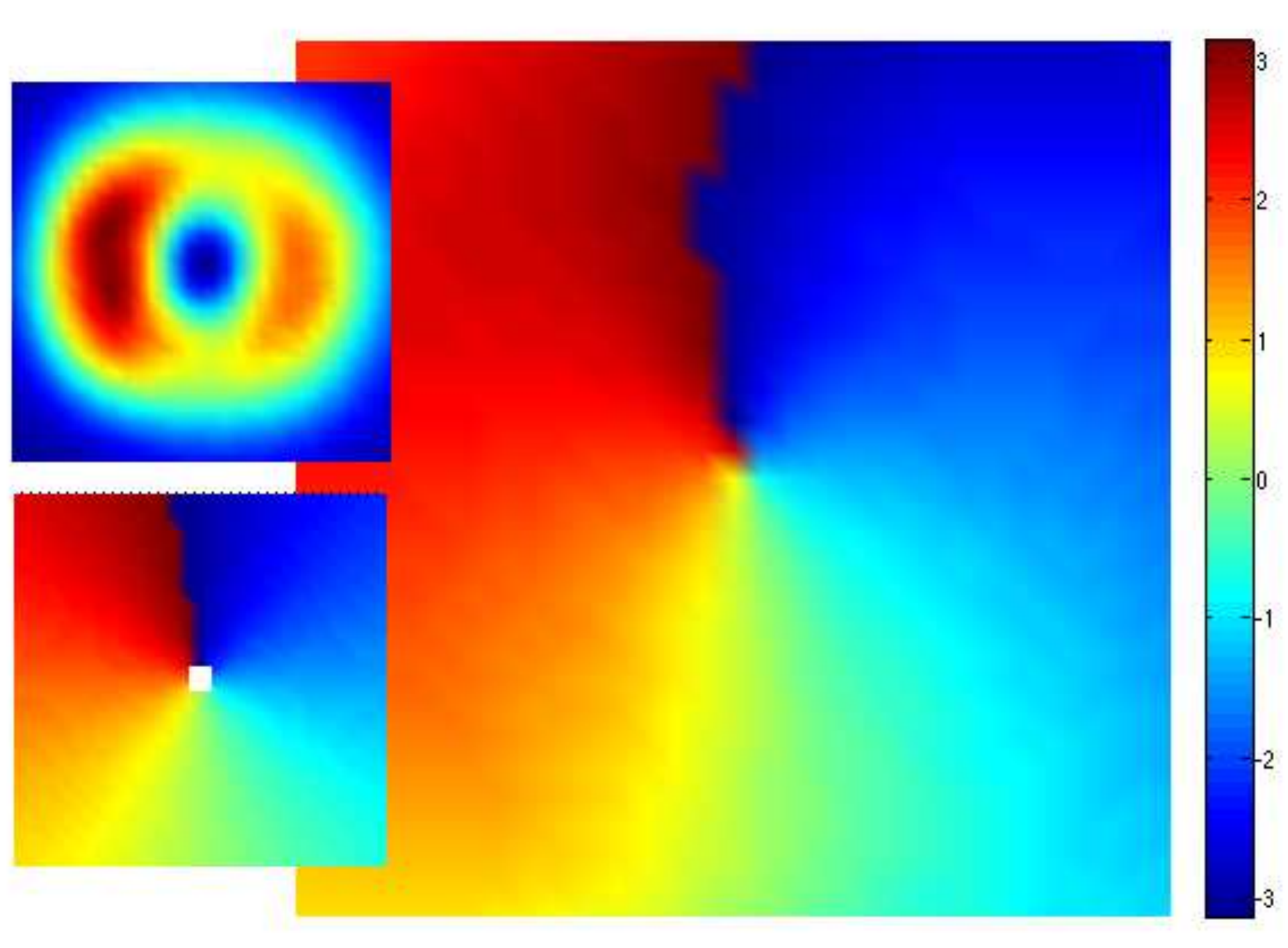}
\caption{Spatial phase of OAM state with $L_z=-1$ that shows characteristic phase wrapping in the clockwise direction. {\it Insets:} (a) Intensity distribution of the mode, (b) theoretical distribution of spatial phase}
\label{fig:oam_minus}
\end{figure}

The results of phase reconstruction of the OAM states are shown in
Figs.~\ref{fig:oam_plus} and~\ref{fig:oam_minus} for $L_z=+1$ and $L_z=-1$
states, correspondingly. The spatial pattern reveals clockwise and
counterclockwise phase wrapping, which uniquely identifies the two states. The
position of the phase discontinuity in Figs.~\ref{fig:oam_plus}
and~\ref{fig:oam_minus} allows direct measurement of the relative phase of
the states.

In summary, OAM states propagating in optical waveguides demonstrate distinct
polarization and spatial patterns. Using polarization and phase-sensitive
C$^2$-imaging methods, I have completely characterized these modes, revealing
their unique polarization and phase singularities, the modal power and also
the relative phase of the states.

\newpage
\section{Resonant mode coupling in leakage channel fibers}
\label{section:LCF}

For an updated description of this experiment, I refer the reader to a recent
preprint ``Resonant Bend Loss in Leakage Channel Fibers'' by R. A. Barankov,
K. Wei, B. Samson, and S. Ramachandran posted at arXiv:1205.5584v1
[physics.optics]. Below is the original write-up as submitted to the library
of Boston University.

In this section, I describe modal characterization of large-mode-area leakage
channel fibers, which demonstrate dramatic power loss at certain coiling
radius. Using C$^2$-imaging, I experimentally attribute this anomaly to
a new physical mechanism of resonant mode-coupling~\cite{Barankov2012_LCF}.

Resonantly enhanced leakage-channel fibers (LCFs) with large mode areas are
designed to provide high-power propagation of diffraction-limited beams in
high-power fiber lasers~\cite{Dong2009}. The microstructure of these fibers is
tailored to enhance the loss of higher-order modes (HOMs) while maintaining
tolerable loss of the fundamental mode, resulting in single-mode operation
with large field diameters. In the existing designs~\cite{Dong2009}, the
boundary between the light-guiding core and the cladding consists of low-index
circular rods, so that index-continuity of the boundary in the azimuthal
direction is broken. As a result, all modes propagating in the fiber are
coupled to the radiative modes of the cladding. However, the presence of gaps
at the interface naturally promotes significant differential power-loss
dominated by HOMs. A careful study of the mechanism and different possible
fiber-designs indicates that high confinement loss for HOMs and low
confinement loss for the fundamental mode can be achieved, leading to
effectively single-mode operation of the structures~\cite{Dong2009}.

The physical mechanism underlying the LCF design suggests sensitivity of light
propagation to the coiling conditions of the fibers. Indeed, it is well-known
that coiled fibers experience power loss resulting from bend-induced coupling
of the guided modes to the radiative modes of the
cladding~\cite{Marcuse1976,Wong2005,Dong2006,Saitoh2011}. In this case, the
loss is significantly higher for HOMs than for the fundamental mode, as
dictated by larger modal overlap of HOMs and the cladding modes. Moreover,
when the cladding modes become quasi-guided in coated
fibers~\cite{Renner1992}, a resonant behavior is observed~\cite{Murakami1978}.

The bend-loss becomes significant for small enough coiling
radii~\cite{Love1989}. The critical radius has been estimated in the case of
photonic-crystal fibers~\cite{Birks1997} as $R_c\sim \Lambda^3/\lambda^2$ in
the short-wavelength limit, where $\lambda$ is the wavelength of light, and
$\Lambda$ is the characteristic core size~\cite{Birks1997,Nielsen2004}. This
estimate should also hold for leakage-channel fibers characterized by similar
geometry. For example, for the LCFs with core size $\Lambda\sim 50\, {\rm\mu
m}$, significant bend loss is expected at $R_c\sim 10 \, {\rm cm}$ in the
$1\, {\rm \mu m}$ spectral range.

In addition to the power-loss resulting from direct radiative coupling of the
core modes, bending of fibers induces inter-modal coupling in the
core~\cite{Russell2006,Love2007}. The coupling increases dramatically when the
effective indices of the two modes approach one another as a function of the
fiber curvature. The crossing occurs at some critical radius, which, in
general, depends on the properties of the coupled modes. Interestingly, the
crossing between the fundamental and the lowest HOM is also expected at the
critical radius $R_c$ corresponding to onset of significant
bend-loss~\cite{Russell2006}.

\begin{figure}[!htb]
\centering
\includegraphics[width=11cm]{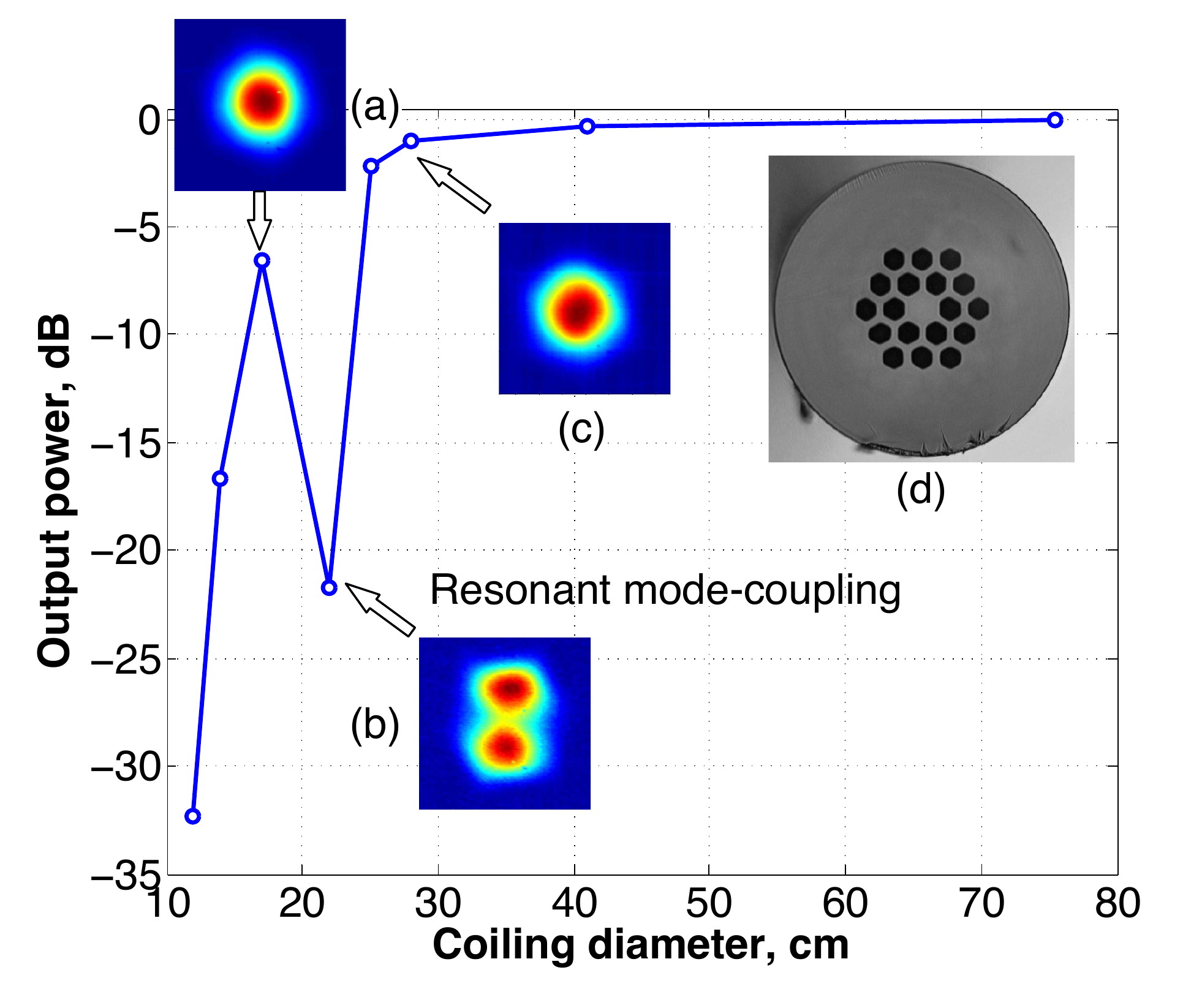}
\caption{Output power as a function of coiling diameter. Insets (a-c): Output mode profiles at different coiling diameters; (d) Cross-section of the LCF}
\label{fig:LCF_power}
\end{figure}

In this work, we explore the effect of resonant mode-coupling on the bend-loss
in a large-mode area LCF and observe an enhanced power loss at a certain
coiling diameter. The results of power measurements in the $1\, {\rm \mu m}$
spectral range are shown in Fig.~\ref{fig:LCF_power}. We observe a dramatic
decrease of the output light power at a specific coiling diameter, while the
power recovers to the levels offset by the usual mechanism of power loss
outside the resonance. At small coiling radii, as expected, one obtains
significant loss of power in accord with the usual bend-loss
mechanism~\cite{Marcuse1976,Birks1997}. We characterize the observed anomaly
by C$^2$-imaging method~\cite{Schimpf2011}. Specifically, we find that for
non-critical coiling radii, light propagation in the tested LCF is dominated
by the fundamental mode with HOM extinction below $-25\,{\rm dB}$. Thus, at
non-critical radii the power-loss can be explained by the usual bend-loss
mechanisms~\cite{Birks1997,Russell2006}. In contrast, at the critical coiling
radius, a higher-order mode dominates propagation, indicating resonant
coupling of the fundamental mode to HOMs, which, by the design of LCFs,
immediately radiate out of the core. Interestingly, this mechanism is
reminiscent of resonant mode coupling observed in coated single-mode
fibers~\cite{Renner1992}. While simple bend-loss measurements shown in
Fig.~\ref{fig:LCF_power} reveal this anomalous behavior, C$^2$-imaging method
allows us to quantify this effect. As a result, this precise characterization
method provides critical feedback for future fiber-designs, in which the
critical radius $R_c$ should be defined for specific amplifier packaging
constraints.

\begin{figure}[!htb]
\centering
\includegraphics[width=11cm]{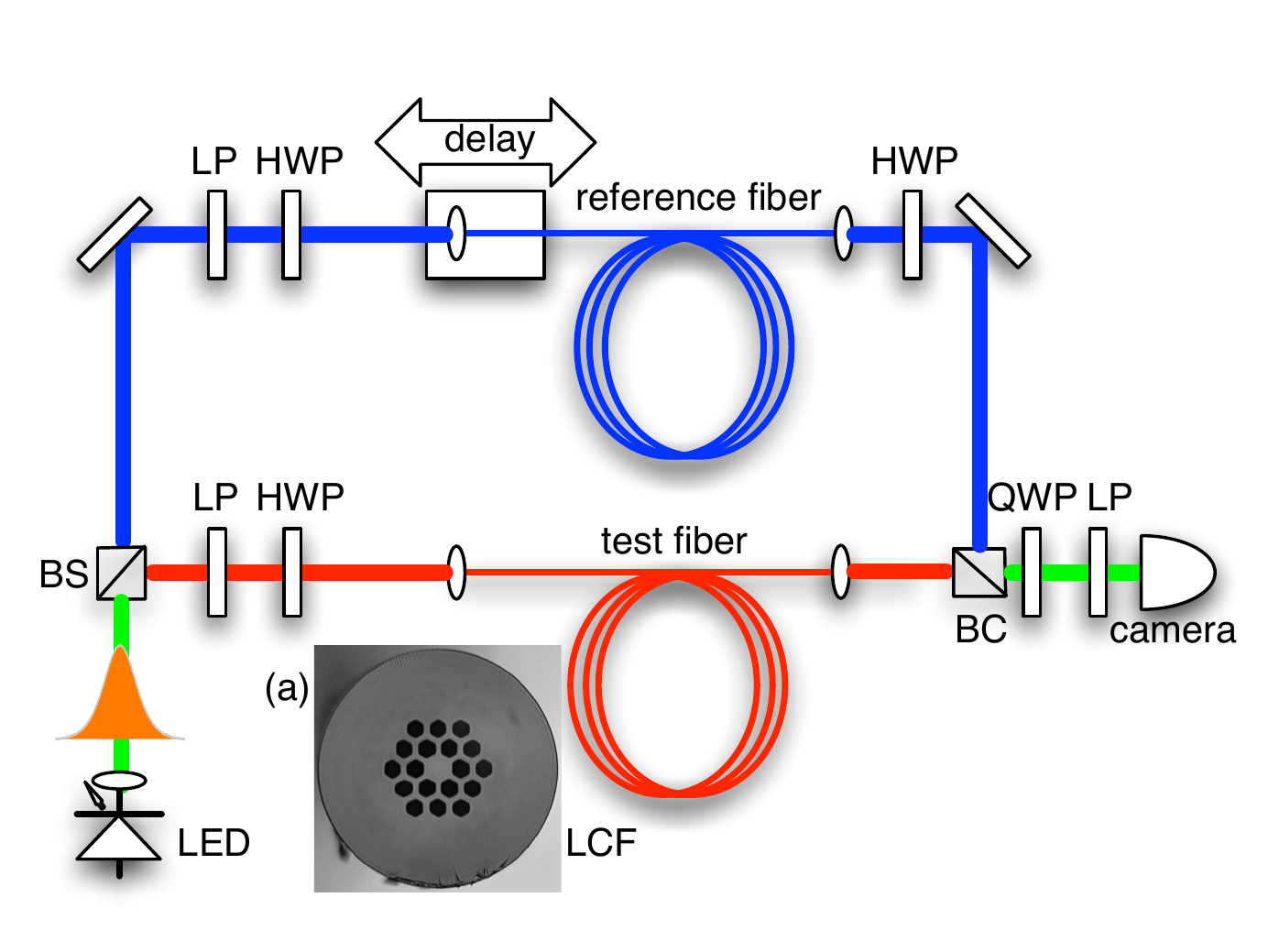}
\caption{C$^2$-imaging setup: LED – light-emitting diode, LP-linear polarizer,
HWP – half-wave plate, QWP – quarter-wave plate, BS – beam-splitter, BC – beam
combiner. Inset (a): Cross-section of the LCF}
\label{fig:LCF_setup}
\end{figure}

The tested LCF of $285\, {\rm cm}$-length has a core diameter of $50\,{\rm\mu
m}$ and cladding diameter of $400\,{\rm\mu m}$. The two rings of low-index
(fluorine-doped) silica regions shown in Fig.~\ref{fig:LCF_setup}(a) provide
the leakage channel. The core, made of silica, is index matched to the outer
silica glass. A high-index regular acrylate coating applied to the cladding
ensures stripping out of the cladding modes. The LCF has been designed to have
negligible HOM content at lengths greater than $3\,{\rm m}$. The input end of
the fiber was spliced to a single-mode fiber to provide same in-coupling
conditions throughout the experiments.

The modal content of the LCF was analyzed using C$^2$-imaging method developed
in Chapter~\ref{chapter:Formalism}. The basic idea of the method is to study
the interference of the beam radiated from an optical waveguide with an
external reference beam, and detect different waveguide modes in the
time-domain by changing the relative optical paths of the two beams.

Figure~\ref{fig:LCF_setup} shows a schematic diagram of the experimental setup
based on a Mach-Zehnder interferometer. We used a superluminescent diode (SLD)
source centered at $\lambda\sim 1050\,{\rm nm}$ with a spectral width of $\sim
30\, {\rm nm}$. The output beam of the LCF (focused at the imaging plane)
interferes with the collimated reference beam radiated from the reference
fiber. The length of polarization-maintaining reference fiber is chosen to
compensate for the optical path difference between the two paths and also to
reduce the effects of group-velocity dispersion of the LCF. The latter is
important since we employ a light source with a relatively broad spectrum,
which, in the absence of dispersion compensation, leads to significant
dispersion broadening of the cross-correlation signal. In particular, the
mode-specific resolution of C$^2$-imaging method is defined by the spectral
width of the source and also by the dispersion mismatch of the reference and
the test modes. In large-mode-area LCFs, the material dispersion dominates the
spectral broadening, which results in similar dispersion properties of HOMs.
We chose the length of the reference fiber to match the material dispersion of
LCF and, thus, reduced the dispersive broadening of all the modes
simultaneously.

The cross-correlation signal ${\cal
P}(\vec r, \tau)$ as a function of the group delay $\tau$ and the coordinate
$\vec r$ in the imaging plane is detected by the camera for different
positions $d=c\tau$ ($c$ is the speed of light in vacuum) of the delay stage:
\be\label{eq:cross_correlation_local}
{\cal P}(\vec r, \tau)=\sum_m p_m{\cal G}_{mr}^2(\tau-\tau_{mr}) I_{m}(\vec r).
\ee
Here, the summation extends over the modes propagating in LCF, $p_m$ is the
relative modal power of $m$-th mode ($\sum_m p_m=1$), ${\cal G}_{mr}(\tau)$ is
the mutual coherence function of the reference and test beams, $I_{m}(\vec
r)$ is the modal intensity, and $\tau_{mr}$ is the relative group delay of the
$m$-th mode with respect to the reference mode. 

Coiling of the fiber affects polarization properties of the test beam. To
extract the power of every elliptically-polarized mode, we have recorded the
cross-correlation trace for two orthogonal polarization states of the
reference beam. The resulting trace, combining the two measurements, is
represented in Eq.~(\ref{eq:cross_correlation_local}). The modal intensity
distribution $I_m(\vec r)$ of every mode was obtained by integrating this
expression over the time extent of the mode. The relative power of the modes
is encoded in the net cross-correlation trace obtained by integration of the
spatially-dependent trace in Eq.~(\ref{eq:cross_correlation_local}) over the
imaging plane position.

\begin{figure}[!htb]
\centering
\includegraphics[width=11cm]{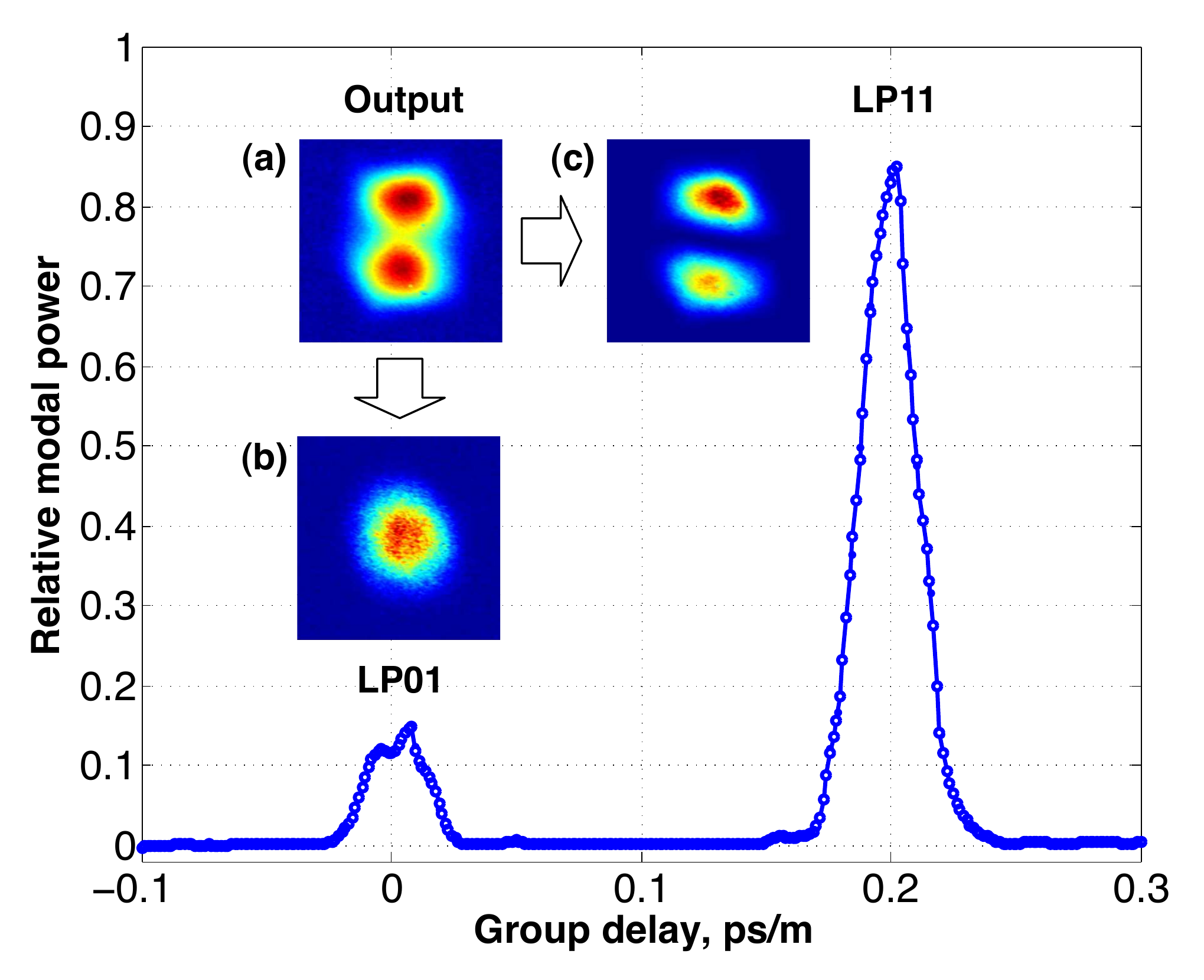}
\caption{Relative modal power as a function of group delay at the resonance (D=22cm). Insets: output image and the images of reconstructed modes}
\label{fig:LCF_resonance}
\end{figure}

The result of this procedure, applied to the cross-correlation traces recorded
at the critical coiling radius, is shown in Fig.~\ref{fig:LCF_resonance}. In
this figure, the cross-correlation peaks identify the modes propagating in the
fiber at the corresponding relative group delays. The shape of the peaks
reflects the corresponding mutual coherence functions, while the peak values
encode the relative power of the modes. The insets demonstrate the intensities
of the reconstructed modes.

The dependence of the output power on the coiling diameter was measured using
a power meter. The results are shown in Fig.~\ref{fig:LCF_power}. We observe a
dramatic decrease of the output light power at a specific coiling diameter.
The observed critical radius $R_{exp}\approx 11 \,{\rm cm}$ is close to the
estimate $R_c\sim 10 \,{\rm cm}$, suggesting resonant mode-coupling as the
mechanism responsible for the anomalous power loss. Indeed, at the resonance,
the output image demonstrates domination of HOM’s as shown in
Fig.~\ref{fig:LCF_power}(b), while the output images in
Fig.~\ref{fig:LCF_power}(a) and~\ref{fig:LCF_power}(c), recorded at coiling
diameters outside the resonance, indicate single-mode operation.

\begin{figure}[!htb]
\centering
\includegraphics[width=11cm]{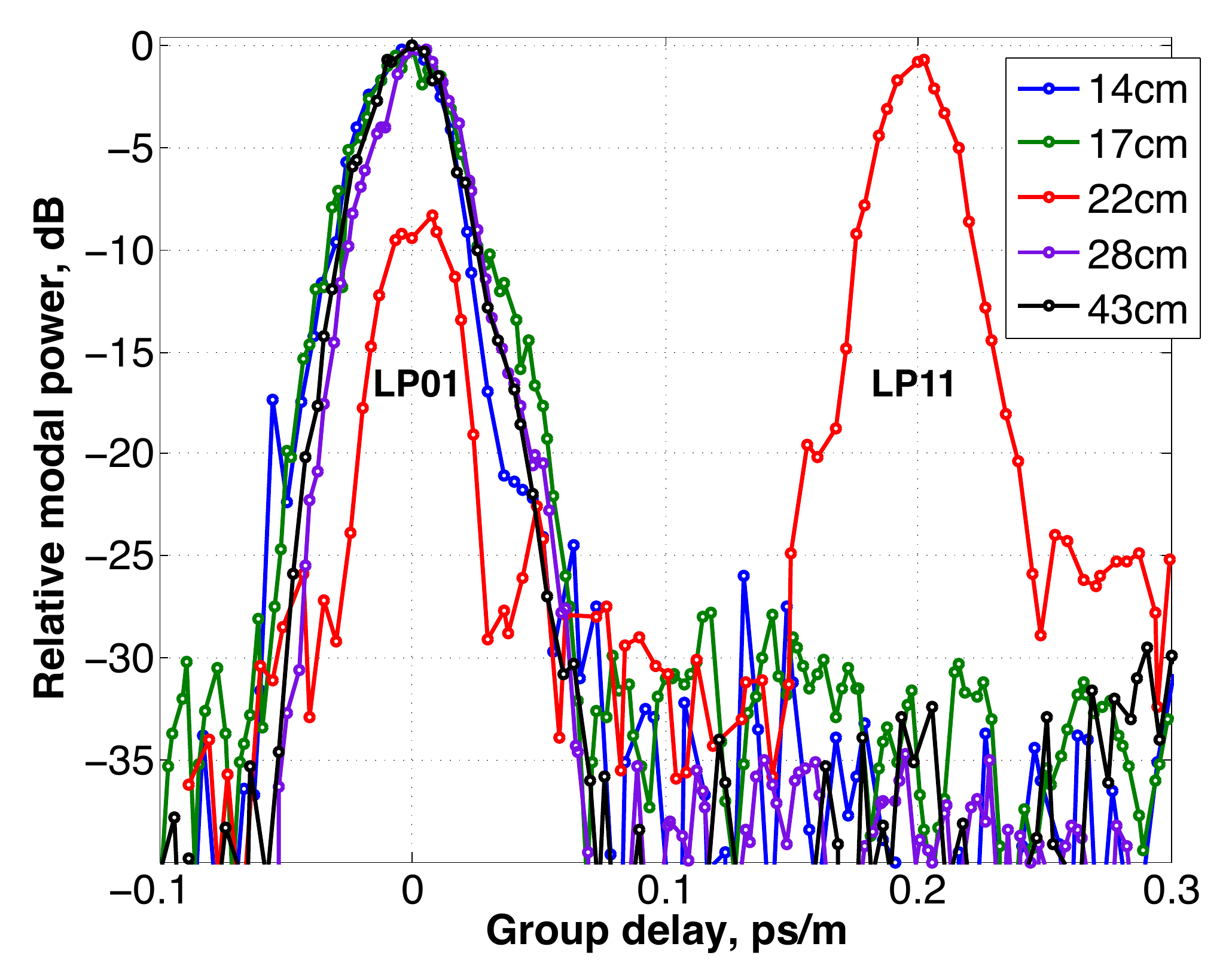}
\caption{Relative modal power of $LP_{01}$ and $LP_{11}$ (peak values) as a function of the relative group delay, for different coiling diameters}
\label{fig:LCF_out_resonance}
\end{figure}

C$^2$-imaging~\cite{Schimpf2011} provides insight into the resonant behavior.
The envelope of the integrated correlation trace shown in
Fig.~\ref{fig:LCF_resonance} at the critical coiling diameter demonstrates two
modes ($LP_{01}$ and $LP_{11}$) propagating with a relative group delay of
about $0.2\, {\rm ps/m}$, with the fundamental mode contributing only about
15\% of the total power. In contrast, at other coiling diameters, HOM’s are
suppressed at the power level below $-25\,{\rm dB}$, as shown in
Fig.~\ref{fig:LCF_out_resonance}. The strength of the resonance depends on the
length of the coiled fiber. In a set of similar measurements conducted on the
LCF of smaller length of about $180\,{\rm cm}$, a relatively shallow resonance
was found at the same coiling diameter. We expect a significantly deeper
resonance for longer fiber lengths.

In summary, coiled few-mode fibers experience bend-induced coupling between
the core modes and power-loss via direct coupling of core modes to the
radiative modes of the cladding. Using C$^2$-imaging, we explore the interplay
of these phenomena in LCFs. We identify a new resonant power-loss mechanism in
these fibers, in which higher-order core modes mediate coupling of the
fundamental mode to the radiative modes. The effect becomes evident at a
specific coiling diameter, where we observe a dramatic decrease of the output
power. Outside the resonance, the power recovers to the levels determined by
the usual bend-loss mechanism. In this work, we quantify this anomaly by
C$^2$-imaging, thus providing critical feedback for the fiber-based amplifier
designs with certain coiling constraints.

\cleardoublepage

\chapter{Conclusions}
\label{chapter:Conclusion}
\thispagestyle{myheadings}

In this work, I have developed a novel interferometric method, called
C$^2$-imaging, suitable for complete characterization of waveguide modes. In
particular, using this method one obtains the modal weights, relative group
delays and dispersion properties of the modes. Modifications of the method,
involving polarization post-selection and the analysis of high-frequency
oscillations of interferometric signal, reveal polarization and phase
distributions of all the modes.

The developed method has been successfully applied to modal characterization
of waveguides with distinct modal properties. In particular, in a specialty
higher-order mode fiber, the method allowed accurate reconstruction of the
modes having small relative group delays and distinct dispersive behavior. The
reconstructed modal weights agree well with the mode-conversion of the
long-period grating spectrum. The multi-path interference values down to
$-30\, {\rm dB}$ can be accurately retrieved. The concept of
dispersion-compensation was demonstrated in the case of large-mode-area
fibers, where sufficient temporal resolution was achieved to observe temporal
birefringent splitting between odd and even higher order modes propagating in
the polarization maintaining fiber.

Polarization-sensitive modification of C$^2$-imaging method was successfully
tested for modal characterization of a specialty fiber supporting several
vector modes. The reconstructed polarization patterns, demonstrating
non-trivial space dependence, compare very well with the theoretical
predictions. In addition, I have directly observed optical orbital angular
momentum states propagating in the fiber, by employing circular polarization
selectivity of the modes. The observed phase patterns of the modes are
consistent with the theoretical expectations for the corresponding spatially
dependent Stokes parameters.

Phase-sensitivity of C$^2$-imaging was also demonstrated in the experiments
with the same specialty fiber. I have directly measured, for the first time to
the best of my knowledge, the phase distribution of the orbital momentum
states, using a phase-retrieval algorithm based upon analysis of
high-frequency oscillations of the interferometric cross-correlation signal.
As a result, I have completely characterized the vortex modes and identified
their relative power and the complex relative phase.

I have also used C$^2$-imaging to explore the resonant behavior of power
output in large-mode-area leakage-channel fibers. In particular, I have
identified the resonant mode-coupling as the mechanism responsible for the
observed dramatic power loss at a specific coiling radius.

The versatility of C$^2$ imaging, demonstrated in this work, identifies this
method as a powerful new tool, suitable for waveguide characterization in
optical applications.

I kindly acknowledge support of this project by ARL Grant No. W911NF-06-2-0040, ONR Grant Nos. N00014-11-1-0133 and N00014-11-1-0098.

\cleardoublepage

\newpage
\singlespace
\bibliographystyle{apalike}

\bibliography{thesis}

\cleardoublepage

\addcontentsline{toc}{chapter}{Curriculum Vitae}

\thispagestyle{empty}

\begin{center}
{\large\bf CURRICULUM VITAE}\\
\vspace{0.2in}
{\large {\bf Roman A. Barankov}}

\medskip

e-mail: barankov@bu.edu; voice: 617-620-3408\\
Laboratory of Nanostructured Fibers \& Nonlinear Optics\\
ECE Department and Photonics Center, Boston University\\
8 Saint Mary's Street, Boston, MA 02215

\end{center}

\indent{\large\bf Education}
\medskip

\begin{tabular}{p{2.cm}p{11.5cm}}

2012 &
{\bf Boston University, Boston, Massachusetts}\\ & -- M.S. in Electrical Engineering (May 2012), \\
 & Research advisor -- Prof. Siddharth Ramachandran\\

&\\

2006 &
{\bf Massachusetts Institute of Technology, Cambridge, Massachusetts}\\ & -- Ph.D. in Physics\\
& Research advisor -- Prof. Leonid Levitov\\

&\\

1998 & {\bf Moscow Engineering Physics Institute, Moscow,
Russia}\\ & -- Diploma with honors in Physics
\end{tabular}

\medskip

\indent{\bf \large Employment}

\medskip

\begin{tabular}{p{2.cm}p{11.5cm}}
2007-2012 & {\bf Boston University,
Boston, Massachusetts}\\ 
& -- Research Assistant at the Department of Electrical and Computer Engineering\\
& -- Research Scientist at the Physics Department\\

&\\

2010-2012 & {\bf Simmons College, Boston, Massachusetts}\\
& -- Lecturer at the Department of Chemistry and Physics\\

&\\

2005-2007 & {\bf University of Illinois at Urbana-Champaign, Urbana, Illinois}\\ 
& -- Postdoctoral Research Associate at the Physics Department\\

&\\

2002-2005 & {\bf Massachusetts Institute of Technology, Cambridge, Massachusetts}\\ 
& -- Research Assistant at the Department of Physics
\end{tabular}

\medskip

{\bf\large Research Area}

\medskip

Applied Optics: interferometric methods for waveguide characterization\\
\indent Theoretical condensed matter and atomic physics:\\
\indent many-body dynamics of cold atoms and quantum fluids

\medskip

{\bf\large Related Professional Experience}

\medskip

Member of the American Physical Society (since 2003)\\ 
\indent Referee for Physical Review Letters and Physical Review A

\newpage

{\large \bf Journal Publications}
\begin{itemize}
	
	\item[1.] \emph{``Cross-correlated ($C^2$) imaging of fiber and waveguide modes''}, D.~N.~Schimpf, R.~A.~Barankov, and S. Ramachandran, Optics Express {\bf 19}, Issue 14, pp. 13008-13019 (2011).

	\item[2.] \emph{``Adiabatic nonlinear probes of one-dimensional Bose
gases''}, C.~De~Grandi, R.~A.~Barankov, and A.~Polkovnikov, Phys. Rev. Lett. {\bf
101}, 230402 (2008).

	\item[3.] \emph{``Optimal non-linear passage through a quantum critical point''}, Roman~Barankov and Anatoli~Polkovnikov, Phys. Rev. Lett. {\bf 101},
076801 (2008).

	\item[4.] \emph{``Phase-Slip Avalanches in the Superflow of $^4$He through
	Arrays of Nanosize Apertures"}, David~Pekker, Roman~Barankov, and Paul~M.~Goldbart, Phys. Rev. Lett. {\bf 98}, 175301 (2007).

	\item[5.] \emph{``Coexistence of Superfluid and Mott Phases of Lattice
	Bosons"}, R.~A.~Barankov, C.~Lannert, and S.~Vishveshwara, Phys. Rev. A {\bf 75}, 063622 (2007).

	\item[6.] \emph{``Synchronization in the BCS Pairing Dynamics as a Critical Phenomenon"}, R.~A.~Barankov and L.~S.~Levitov, Phys. Rev. Lett. {\bf 96}, 230403 (2006).

	\item[7.] \emph{``Dynamical Selection in Emergent Fermionic Pairing"},
	R.~A.~Barankov and L.~S.~Levitov, Phys. Rev. A {\bf 73}, 033614 (2006).

	\item[8.] \emph{``Atom-molecule Coexistence and Collective Dynamics near
	a Feshbach Resonance of Cold Fermions"}, R.~A.~Barankov and L.~S.~Levitov,
	Phys. Rev. Lett. {\bf 93}, 130403 (2004).

	\item[9.] \emph{``Solitons and Rabi Oscillations in a Time-Dependent BCS
Pairing Problem"}, R.~A.~Barankov, L.~S.~Levitov and B.~Z.~Spivak, Phys. Rev.
Lett. {\bf 93}, 160401 (2004).

	\item[10.] \emph{``Dissipative Dynamics of a Josephson Junction in the
	Bose-Gases"}, \\ R.~A.~Barankov and S.~N.~Burmistrov, Phys. Rev. A {\bf 67},
	013611 (2003).

	\item[11.] \emph{``Boundary of Two Mixed Bose-Einstein Condensates"},
	R.~A.~Barankov, Phys. Rev. A {\bf 66}, 013612 (2002).
	
\end{itemize}

\end{document}